\RequirePackage{booktabs}
\documentclass[sn-apa]{sn-jnl}

\usepackage{graphicx}%
\usepackage{makecell, multirow, tabularx}
\usepackage{amsmath,amssymb,amsfonts}%
\usepackage{amsthm}%
\usepackage{mathrsfs}%
\usepackage[title]{appendix}%
\usepackage{xcolor}%
\usepackage{textcomp}%
\usepackage{manyfoot}%
\usepackage{algorithm}%
\usepackage{algorithmicx}%
\usepackage{algpseudocode}%
\usepackage{listings}%
\usepackage[most]{tcolorbox}
\usepackage{balance}
\usepackage{subcaption}

\usepackage[verbatim]{lstfiracode}
\definecolor{halfgray}{HTML}{2B91AF}
\definecolor{j_frame}{HTML}{D3D4D5}
\definecolor{j_bg}{HTML}{FFFFFE}
\definecolor{j_red}{HTML}{699E08}
\definecolor{j_green}{HTML}{F9025A}
\definecolor{j_cyan}{HTML}{867B67}
\definecolor{j_purple}{RGB}{170, 34, 255}

\usepackage{listings}
\lstdefinelanguage[]{iJava}[]{java}{
    xleftmargin={0.75cm},
    sensitive=true,%
    commentstyle=\color{j_cyan}\ttfamily,
    stringstyle=\color{j_red}\ttfamily,
    keepspaces=true,
    showspaces=false,
    showstringspaces=false,
    rulecolor=\color{j_frame},
    frame=single,
    frameround={t}{t}{t}{t},
    framexleftmargin=6mm,
    numbers=left,
    numberstyle=\tiny\color{halfgray},
    backgroundcolor=\color{j_bg},
    style=FiraCodeStyle,
    basicstyle=\scriptsize\ttfamily,
    keywordstyle=\color{j_green}\ttfamily,
}
\begin{document}

\title[\textbf{\textit{CodeSAM}}: SCRL by Infusing Self-Attention with MCV Graphs]{\textit{CodeSAM}: Source Code Representation Learning by Infusing Self-Attention with Multi-Code-View Graphs}

\author*[3]{\fnm{Alex} \sur{Mathai}}\email{am6215@columbia.edu}
\equalcont{These authors contributed equally to this work.}

\author[2]{\fnm{Kranthi} \sur{Sedamaki}}\email{skranthi4444@gmail.com}
\equalcont{These authors contributed equally to this work.}

\author[4]{\fnm{Debeshee} \sur{Das}}\email{debdas@student.ethz.ch}
\equalcont{These authors contributed equally to this work.}

\author*[5]{\fnm{Noble Saji} \sur{Mathews}}\email{noblesaji.mathews@uwaterloo.ca}
\equalcont{These authors contributed equally to this work.}

\author[1]{\fnm{Srikanth} \sur{Tamilselvam}}\email{srikanth.tamilselvam@in.ibm.com}
\author[2]{\fnm{Sridhar} \sur{Chimalakonda}}\email{ch@iittp.ac.in}

\author[1]{\fnm{Atul} \sur{Kumar}}\email{kumar.atul@in.ibm.com}

\affil*[1]{\orgdiv{IBM Research}, \orgaddress{\country{India}}}
\affil[2]{\orgdiv{IIT Tirupati}, \orgaddress{\country{India}}}
\affil[3]{\orgdiv{Columbia University}, \orgaddress{\country{USA}}}
\affil[4]{\orgdiv{ETH Zurich}, \orgaddress{\country{Switzerland}}}
\affil[5]{\orgdiv{University of Waterloo}, \orgaddress{\country{Canada}}}


\abstract{Machine Learning (ML) for software engineering (SE) has gained prominence due to its ability to significantly enhance the performance of various SE applications. This progress is largely attributed to the development of generalizable source code representations that effectively capture the syntactic and semantic characteristics of code. In recent years, pre-trained transformer-based models, inspired by natural language processing (NLP), have shown remarkable success in SE tasks. However, source code contains structural and semantic properties embedded within its grammar, which can be extracted from structured code-views like the Abstract Syntax Tree (AST), Data-Flow Graph (DFG), and Control-Flow Graph (CFG). These code-views can complement NLP techniques, further improving SE tasks. Unfortunately, there are no flexible frameworks to infuse arbitrary code-views into existing transformer-based models effectively. Therefore, in this work, we propose \textit{CodeSAM}, a novel scalable framework to infuse multiple code-views into transformer-based models by creating self-attention masks. We use \textit{CodeSAM} to fine-tune a small language model (SLM) like CodeBERT on the downstream SE tasks of semantic code search, code clone detection, and program classification. Experimental results show that by using this technique, we improve downstream performance when compared to SLMs like GraphCodeBERT and CodeBERT on all three tasks by utilizing individual code-views or a combination of code-views during fine-tuning. We believe that these results are indicative that techniques like \textit{CodeSAM} can help create compact yet performant code SLMs that fit in resource constrained settings.}




\maketitle

\section{Introduction}

Machine learning for software engineering (ML4SE) is an active area of research that has demonstrated substantial improvements in many SE tasks such as source code completion \citep{xu2022systematic}, automatic code summarization \citep{zhang2022survey} and semantic code search \citep{guo2020graphcodebert}. The success of these ML4SE pipelines largely depends on effectively learning source code representations. Hence, introducing novel ML techniques that incorporate information from source code is an active area of research \citep{10.1145/3212695}. With the advent of language models, many works have explored methods to adapt natural language techniques to source code directly \citep{feng2020codebert, nijkamp2022codegen, xu2022systematic}. Although popular NLP techniques show impressive results when applied to source code \citep{feng2020codebert}, various studies emphasize that source code is a richer construct and should not be treated simply as a collection of tokens or natural language text \citep{zhang2019novel}. 
As a result, contemporary research in ML4SE deliberately avoids restricting itself to traditional NLP approaches alone. Instead, it strives to leverage program analysis techniques and code-views such as the AST, CFG, and DFG, and integrate them into language models to achieve a more comprehensive understanding of source code.
This can be seen from transformer-based \citep{vaswani2017attention}  SLMs like GraphCodeBERT  \citep{guo2020graphcodebert} and SemanticCodeBERT \citep{du2023pre} that are performant in multiple downstream SE tasks. To accommodate graph-based code-views into a transformer model, works like GraphCodeBERT convert a code-view into a sequence and then append this sequence to the original code tokens. However, upon closer examination, we identify and elaborate on two major drawbacks of this approach - (1) the loss of structural information when flattening graphs and (2) the inability to scale this approach in a multi-code-view setting. 

\begin{enumerate}
    \item \textbf{Flattening Graphs}: Since most code-views are predominantly graph structures, they cannot be directly used by models that consume sequential tokens (like transformers). As a result, most methods like GraphCodeBERT flatten the graph code-views by applying post-processing techniques like extracting all the graph's nodes into a sequence. Flattening graphs into sequences erodes the structural information contained in the edges of the graph code-view. To compensate for this loss, GraphCodeBERT attempts to learn the graph topology using a pre-training task of edge-prediction between the graph nodes. However, we note that it is important to prioritize an approach that takes the complete graph topology as input rather than focusing on mitigating information loss from graph flattening.    
    \item \textbf{Ballooning Input Size}: After the graph is flattened into a sequence, approaches like GraphCodeBERT proceed to append this sequence to the original code snippet tokens. 
    As most SLMs can typically handle sequence lengths of up to $1024$ or $2048$ tokens \citep{beltagy2020longformer}, the original code tokens are often truncated to accommodate the additional code-view sequence. This problem is further exaggerated when considering multiple code-views. Trying to accommodate the flattened sequences of two or more code-views would result in input overflow or extreme truncation of the original code tokens, resulting in the loss of crucial context. Hence, there is a need for a scalable method that incorporates code-views without increasing sequence lengths.
\end{enumerate}
Based on the above observations, we believe that existing methods fall short when it comes to effectively integrating multi-code-view representations into transformer models.
To solve the above drawbacks, we propose \textbf{\emph{CodeSAM}}, a novel approach to infuse information from multi-code-views by modifying the self-attention mechanism for transformers without altering the input token sequence. Additionally, after generating this self-attention mask, we propose two masking strategies for transformer models, that help incorporate local and global context in the source code representations. We also design \textit{CodeSAM} to be code-view agnostic and applicable to any single or multi-code-view graph representation. Hence, we believe that \textit{CodeSAM} is the first work to showcase how to execute a practical and complete pipeline - beginning from the static analysis of non-compilable code snippets, abstracting this analysis into a language and code-view agnostic format and finally integrating this format into transformer models without impacting input sequence lengths. This is unlike other works that implement methodologies that are specific to one code-view or methods that incorporate multiple code-views by significantly increasing input sequence length like GraphCodeBERT and SemanticCodeBERT.
\newline
\newline
\fbox{\begin{minipage}{\hsize}
Our core contribution is a novel, code-view agnostic, and flexible framework that \textit{infuses} multi-code-view information via self-attention masking strategies into transformer-based models.
\end{minipage}}
\newline
\newline

\textit{\textbf{Paper Organization:}} We elaborate on \textit{CodeSAM}, our novel methodology used for source code representation learning in Section \ref{sec:graph_rep}. In Section \ref{sec:codeviewgeneration}, we present a systematic analysis of the most relevant code-view generating tools often required for ML pipelines, and describe how we arrived at an appropriate tool to be plugged into our \textit{CodeSAM} pipeline. Since there exist many static analysis tools that are language specific  \citep{kovalenko2019pathminer,bieber2022library}, or code-view specific, without support for customizing or combining multiple code-views, we believe that this section will help future ML4SE researchers in selecting appropriate analysis tools. Section \ref{sec:experiments} deals with the evaluation of our methodology through experiments on \textit{code clone detection}, \textit{semantic code search}, and \textit{program classification}. Section \ref{sec:related_work} contains the relevant related literature. Section \ref{sec:threats} entails the Threats to Validity of this work and is followed by Conclusion and Future Work in Section \ref{sec:conclusion}.

\section{Code-View Infused learning}\label{sec:graph_rep}

In this section, we introduce and explain \textit{CodeSAM} - our approach to infuse multiple code-views without flattening its graph structure and simultaneously leaving the input token sequence untouched. We achieve this by choosing to carefully modify the model's attention mechanism  based on the given custom (or combined) code-view. As this approach is essentially a masking procedure, it is applicable to any self-attention based model that consumes sequential data. 


\begin{figure}
    \centering
    \includegraphics[width=0.3\textwidth]{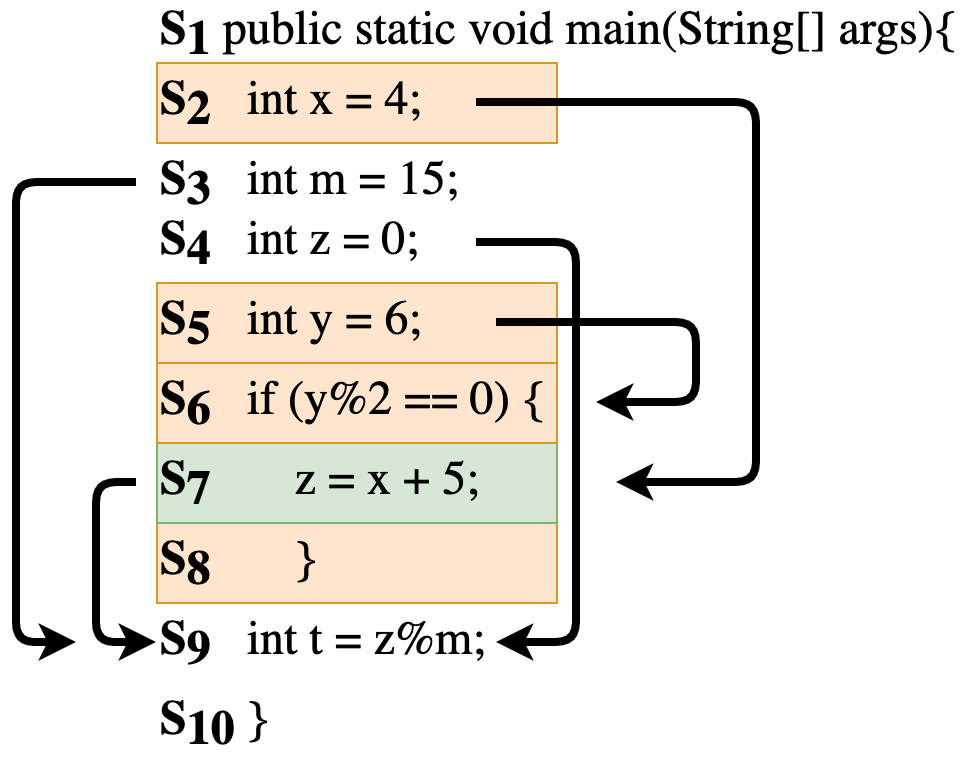}
    \caption{Example $S'_7$ where $\mathbb{T}_{X} = \mathbb{T}_{A+D}$ (Data flow in Black Arrows)}
    \label{fig:highlight_code}
\end{figure}   
\begin{figure}
    \centering
    \includegraphics[width=0.65\textwidth]{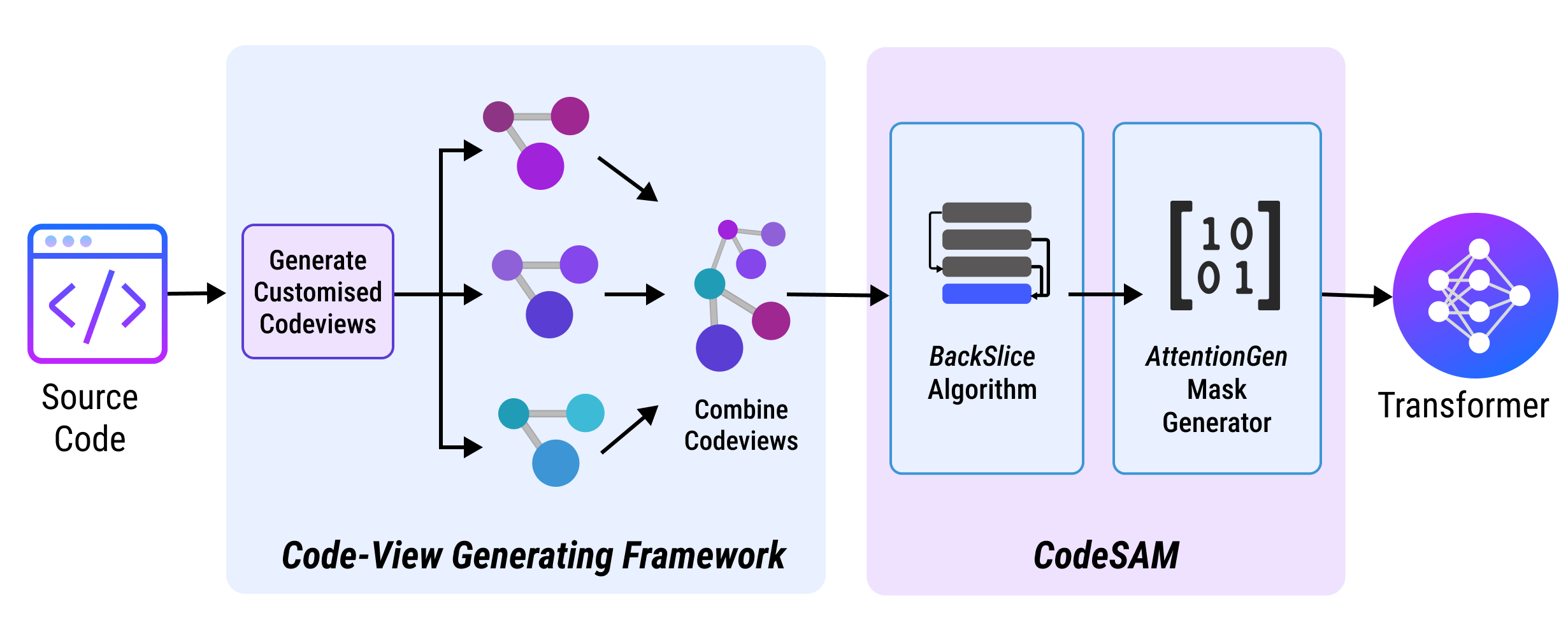}
    \caption{End-to-End System to infuse code-view information into Transformer models}
    \label{fig:system}
\end{figure} 

The self-attention mechanism is an operation that computes a similarity measure between every pair of input tokens. During the course of training, the self-attention mechanism learns an attention matrix $A \in \mathbb{R}^{N \times N}$ where $N$ is the number of input tokens. The matrix $A$ can thus be viewed as a graph adjacency matrix connecting all tokens in the input. With this view in mind, it is possible to use insights from the highly structured graph-based code-views, to modify $A$ and influence the model's learning process. In this paper, we introduce \textit{CodeSAM}  that takes a code-view as input and translates it into an attention mask $A' \in \{0,1\}^{N \times N}$. By replacing $A$ with the Hadamard product $A' \odot A $ in the self-attention operation, we can force each token in the input sequence to attend to only those tokens in the code that are connected to it by a code-view. $A'$ can thus be used to modify $A$ by infusing rich semantic information extracted from the code-views. It is important to note that our approach has no effect on the input sequence to the model, whereas, adding sequences of flattened code-views dramatically increases input length. This is particularly advantageous for fixed (or short) length models like SLMs, where we often need to truncate the original code snippet to accommodate the flattened code-views.
 Additionally, \textit{CodeSAM} can be easily applied to any code-view that is needed for specific downstream tasks. 
 Our implemented pipeline can take any code-view as input, process this code-view to create sentence-level masks and subsequently translate this to the attention mask $A'$ (Fig. \ref{fig:system}). In the following sections, we explain (i) \textit{BackSlice} (Algorithm \ref{alg:backwardslice}) - a backward slicing algorithm that creates sentence level masks and (ii) \textit{AttentionGen} - a method that converts these masks to the attention mask ($A'$).

(i) \emph{\textbf{BackSlice.}} To better understand \emph{BackSlice}, we first introduce a few notations. We represent a code snippet ($CS$) as a list of statements - $\{S_1,S_2,\cdots,S_M\}$ where $M$ is the total number of lines. We refer to the generated code-views using the following naming convention - we denote the AST as $\mathbb{T}_{A}$, the statement level DFG as $\mathbb{T}_{D}$ and the statement level CFG as $\mathbb{T}_{C}$. A composite graph, which is a combination of views will have the $+$ operator in the subscript. For example, $\mathbb{T}_{A+D}$ is the combination of $\mathbb{T}_{A}$ and $\mathbb{T}_{D}$. 


\begin{algorithm}[t]
  \caption{\textbf{\emph{BackSlice}}}
  \label{alg:backwardslice}
  \begin{algorithmic}[1]
    \Require : $S_m$ (current statement); $\mathbb{T}_{X}$ (code-view); $S_{holders}$ (grammar statement holders)
    \State $S'_m \gets \varnothing$
    \State $visitedSet$ $\gets \{S_m\} $
    \State $canVisitSet$ $\gets \{S_m\} $
    \While{ $canVisitSet$ $\neq \varnothing$ }
            \State \% Take an element from the $canVisitSet$ \%
            \State $node$ $ \gets $ pop($canVisitSet$)
            \If{ $node \notin S_{holders}$ }
                \State $S'_m \gets S'_m$ $\cup$ $\{node\}$
            \EndIf
            \State \% Get parents of $node$ from $\mathbb{T}_{X}$ \%
            \State parents $\gets $ getParents($node$,$\mathbb{T}_{X}$)
            \For{$par \in$ parents}
                \If{$par \notin$ $visitedSet$ }
                    \State $canVisitSet$ $\gets$ $canVisitSet$ $\cup$ $\{par\}$
                    \State $visitedSet$ $\gets$ $visitedSet$ $\cup$ $\{par\}$
                \EndIf
            \EndFor
    \EndWhile
    \Return{$S'_m$}
  \end{algorithmic}
\end{algorithm}

We now discuss the three inputs required to run the \emph{BackSlice} algorithm. First, we provide $S_m$ - the statement that we want to analyze. Second, we pass $S_{holders}$ - the list of language-specific grammar constructs that act as groupings of statements. For example, in Java, the construct $``block" \in S_{holders}$. Our third and final argument is $\mathbb{T}_{X}$ which is a code-view generated by any static analysis tool. With these inputs, for each statement $S_{m} \in CS $, identified through $\mathbb{T}_{C}$, we perform \emph{BackSlice} to isolate those code snippet lines that are relevant to $S_{m}$ with respect to code-view $\mathbb{T}_{X}$. Hence for every $S_{m}$,  we derive its corresponding sentence mask ($S'_{m}$) which is a set that contains important statements related to $S_{m}$. Thus $S'_{m} = \{S_{m1},S_{m2},...,S_{mK} \}$ where $K$ is $|S'_m|$ and every $S_{mk} \in CS$. In Fig. \ref{fig:highlight_code}, we highlight line $S_7$ (in green) and its corresponding sentence mask $S'_7$ (in brown). Here $S'_7$ is the set $\{S_2,S_5,S_6,S_7,S_8\}$. As shown in Algorithm \ref{alg:backwardslice}, to compute such a mask, we first start with three sets - $S'_m$, $visitedSet$ and $canVisitSet$ (Lines $1-3$). The $visitedSet$ keeps track of visited statements and the $canVisitSet$ keeps track of the statements that will eventually be visited. As long as the $canVisitSet$ is not empty, we perform two operations. Firstly, we pop out an element (referred as $node$) from the $canVisitSet$ and add it to $S'_m$ only if it is a statement node (Lines $6-9$). This additional check weeds out AST specific constructs  $ \in  S_{holders}$. Secondly, we use $\mathbb{T}_{X}$ to extract the parents of $node$ (parents have outgoing edges into $node$) and add unseen parents to the $canVisitSet$ and $visitedSet$. By doing so, we connect $S_m$ to only those statements related to it by the code-view $\mathbb{T}_{X}$ (Lines $11-17$). In Fig. \ref{fig:highlight_code}, if we only consider data flow (i.e. $\mathbb{T}_{X} = \mathbb{T}_{D}$), then $S'_7$ would be $\{S_2,S_7\}$. If we only consider structural information (i.e. $\mathbb{T}_{X} = \mathbb{T}_{A}$), then $S'_7$ would be $\{S_6,S_7,S_8\}$. However, if we consider both  (i.e. $\mathbb{T}_{X} = \mathbb{T}_{A+D}$) then $S'_7$ would be the larger set $\{S_2,S_5,S_6,S_7,S_8\}$. 
For every $S_m$ we analyze, we believe that its corresponding mask ($S'_m$) is a good approximation for its surrounding context. Our next step in the process is to use these masks to create the attention mask $A'$.

(ii) \emph{\textbf{AttentionGen.}} In this step, we map every statement ($S_m$) to its corresponding list of code tokens ($tokens_m$). Thus $tokens_m = \{ tok_{m1}, tok_{m2}, ...., tok_{mR} \}$ where $R$ is the number of tokens in $S_m$. Using this mapping, we then derive $tokens'_m$ - the list of important tokens for the mask ($S'_m$). This is done by collecting the list $tokens_j$ for each sentence $S_j \in S'_m$. Thus $tokens'_m = \{ tokens_j \mid \forall S_j \in S'_m \} $. Using the above information, we can construct the attention mask ($A'$) as shown in Equation \ref{eq:attention}. Note that $A' \in \{0,1\}^{N \times N}$ where $N$ is the number of tokens in $CS$.  
\begin{equation} \label{eq:attention}
    \underset{ i,j \in \{1,\cdots,N\} }{A'(i,j)}=
\begin{cases}
1, \; \text{if\;} tok_i \in tokens_m \And tok_j \in tokens'_m \\ 
0, \; \text{Otherwise}
\end{cases}
\end{equation}
From Equation \ref{eq:attention}, it is clear that we allow attention between the token $tok_i$ (in sentence $S_m$) and the token $tok_j$ only if $tok_j$ belongs to a sentence in the mask $S'_m$.  This completes the creation of our code-view inspired attention mask. Fig. \ref{fig:system}, shows an overview of our entire system - starting with a static analysis tool that creates a code-view, the \textit{BackSlice} algorithm that converts the code-view into sentence masks and finally, the \textit{AttentionGen} step that translates the sentence mask to the attention matrix $A'$. In the following section, we elaborate on how we arrived at a static analysis tool and then run the entire end-to-end system on three datasets and record our experimental results.


\label{framework}

\section{Code-View Generating Framework}
\label{sec:codeviewgeneration}

Various works involving ML for code require generating different code-view graphs such as the AST \citep{alon2019code2vec, alon2020structural,  kim2021code, wang2021code, zhang2019novel, zügner2021languageagnostic}, CFG \citep{bieber2020learning, bieber2022static, defreez2018pathbased}, DFG \citep{guo2020graphcodebert, kaufman2021learned, gao2023code}, etc. Some also rely on graphs that combine information from various code-views \citep{vasudevan2021learning, du2023pre}. 
Hence, static analysis tools for code-view generation are popularly used by researchers in ML4SE. However, with such a plethora of tools available, it is crucial to identify the most suitable one for the specific ML dataset and approach at hand.
In what follows, we elaborate on how we compare different tools and then select our tool of choice, which we integrate into the \textit{CodeSAM} pipeline (Fig. \ref{fig:system}). 

\subsection{Satisfying the Code-View requirements of \textit{CodeSAM}}
In this section, we identify and discuss some common characteristics of widely used ML datasets. Based on these characteristics we list down certain requirements that the chosen static analysis tool should ideally support. 

\subsubsection{Tool Support for partial/non-compilable code} 
Typically, ML approaches for various SE tasks such as \textit{semantic code search, code clone detection}, and \textit{program classification}, rely on large datasets consisting of partial/non-compilable code snippets. Often, these large datasets are method-level datasets \citep{codesearchnet, svajlenko2014towards} consisting of individual methods with incomplete information about the source of some external variables and functions. Even in the case of file-level datasets \citep{ibmcodenet}, as each file-level data point is analysed outside of a project, it may have external dependencies which are not resolvable. Hence, regular static analysis tools that need the code to be compilable cannot be used for analyzing such ML datasets. For example, SOOT \citep{soot}, one of the most popular static analysis tools for Java, needs the Java source code to be compilable and for all definitions to be available because it compiles and converts the code into an intermediate representation before performing analysis. Hence, this approach does not work for most popular ML datasets, including the ones we use for our experiments. Thus, the chosen tool must perform static analysis directly on the source code by using reasonable approximations where accurate analysis isn't possible. We discuss on three open source tools we found that work in this setting - (i) \textit{COMEX} \citep{comex}, (ii) \textit{Joern} \citep{yamaguchi2014modeling} and (iii) \textit{python-graphs} \citep{bieber2022library}. As we experiment solely on Java datasets, \textit{python-graphs} is not relevant to the experiments in this paper. Additionally, we note that \textit{COMEX} is a generic tool aimed at generating customizable combined code-view representations, while \textit{Joern} is a specialized tool primarily designed to aid in vulnerability detection.

\subsubsection{Tool Support for inter-procedural analysis} 
ML datasets can also include file-level data points \citep{ibmcodenet}.  Thus, the chosen tool should support both inter-procedural (file-level) and intra-procedural (method-level) analysis. Upon evaluating \textit{COMEX}, we find that it supports out-of-the-box inter-procedural analysis using approximations to alias analysis. In comparison, Joern has limited support for inter-procedural analysis, forcing previous works to re-implement many program-analysis components 
\citep{li2019comparative, pewny2016evilcoder}.

\subsubsection{Tool Support for code-view combinations} 
In order to perform an exhaustive analysis, the chosen tool should be able to generate, customize and combine various code-views with ease. 
This is non-trivial since the tool would have to output each code-view in a generalised format, making it possible to identify and combine common nodes across these representations.
As a result it is possible to find many tools that focus on generating a single code-view, but it is difficult to find tools that have support for generating and combining multiple code-views. For example, \textit{Astminer}\citep{kovalenko2019pathminer} (previously PathMiner) can generate an AST and a submodule in GraphCodeBERT can generate a DFG, but neither can generate combined code-views.  \textit{COMEX} and \textit{Joern}, on the other hand, can provide multiple code-views. \textit{COMEX} can generate any customized combination of code-views from AST, CFG and DFG. \textit{Joern} allows Code Property Graph (CPG) generation from a wide range of languages and implements its own query language to work with the same. The CPG is a combination of the control dependence graph (CDG), data dependence graph (DDG) and the AST.  Despite allowing extraction of individual code-views, \textit{Joern} currently lacks out-of-the-box support for easily extracting arbitrary combinations like ``CDG and AST" or ``DDG and AST". If such combinations are required, the user has to filter out the required information from the CPG.

In addition to the above requirements, we note that it is relatively simpler to implement the \textit{CodeSAM} algorithm when the static analysis tool provides control-flow and data-flow edges between statements rather than producing it at a token level.
Contrasting between \textit{COMEX} and \textit{Joern}, \textit{COMEX} has a statement-level output, whereas \textit{Joern} outputs data-flow edges between AST leaf-nodes and control-flow edges between intermediate AST nodes. 

Based on the above observations, we integrate \textit{COMEX} into the \textit{CodeSAM} pipeline as it provides the most flexibility and the most ease of use. Additionally, \textit{COMEX} is built on \emph{tree-sitter}\footnote{https://tree-sitter.github.io/tree-sitter/} - 
an incremental parser that can parse syntactically incomplete code, and provides a uniform parsing interface for over $40$ programming languages. Thus, as and when \textit{COMEX} is extended to support a new language, this also increases the coverage for the \textit{CodeSAM} pipeline. As of today, \textit{COMEX}  provides both method-level and file-level support for Java and C\#.

Other than the tools mentioned in this section, we also compare \textit{COMEX} against GraphCodeBERT's \citep{guo2020graphcodebert} program analysis module for the sake of completeness. It is important to note that, unlike \textit{COMEX} and \textit{Joern}, the GraphCodeBERT module is not a full-fledged tool. Rather, it is a collection of scripts that compute the data-flow graph for six programming languages. We describe a detailed comparison between \textit{COMEX} and this module in Section \ref{sec:gcb-compare}.




\subsection{Comparison with GraphCodeBERT's analysis}
\label{sec:gcb-compare}

\begin{figure}
    \centering
    \includegraphics[width=0.8\textwidth]{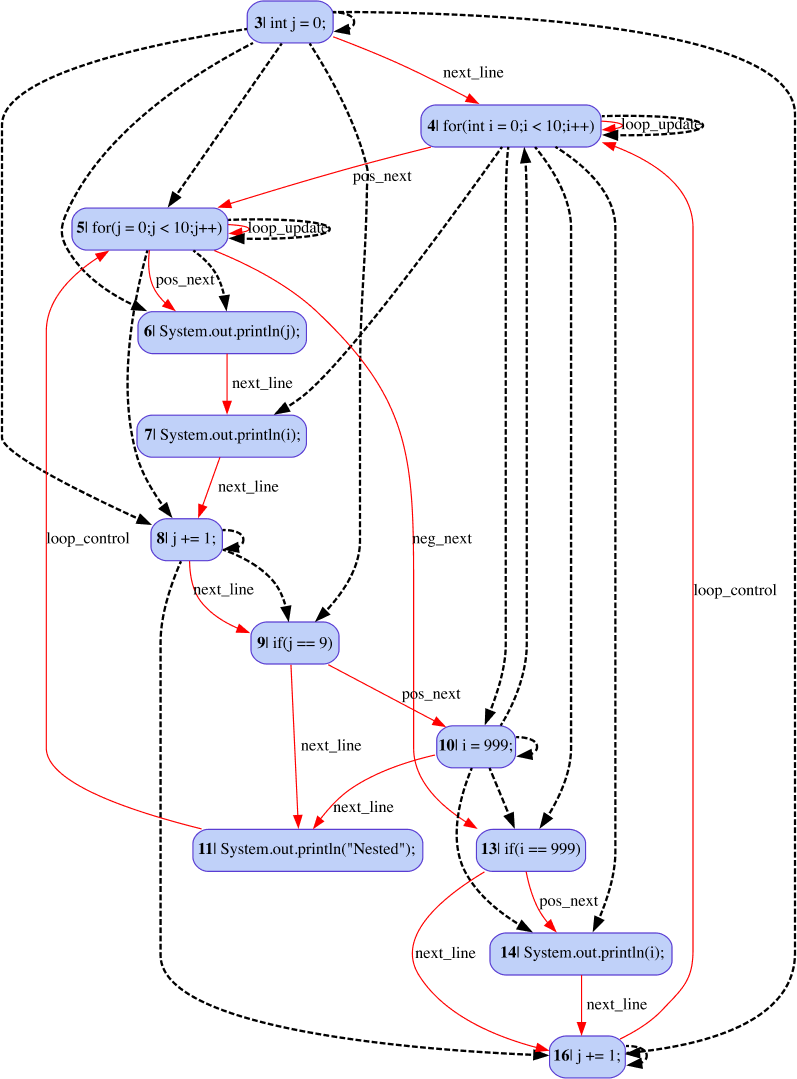}
    \caption{Example of CFG+DFG without RDA (Obtained using GraphCodeBERT's Dataflow logic)}
    \label{fig:dfg_gcb}
\end{figure}   
\hfil
\begin{figure}
    \centering
    \includegraphics[width=0.8\textwidth]{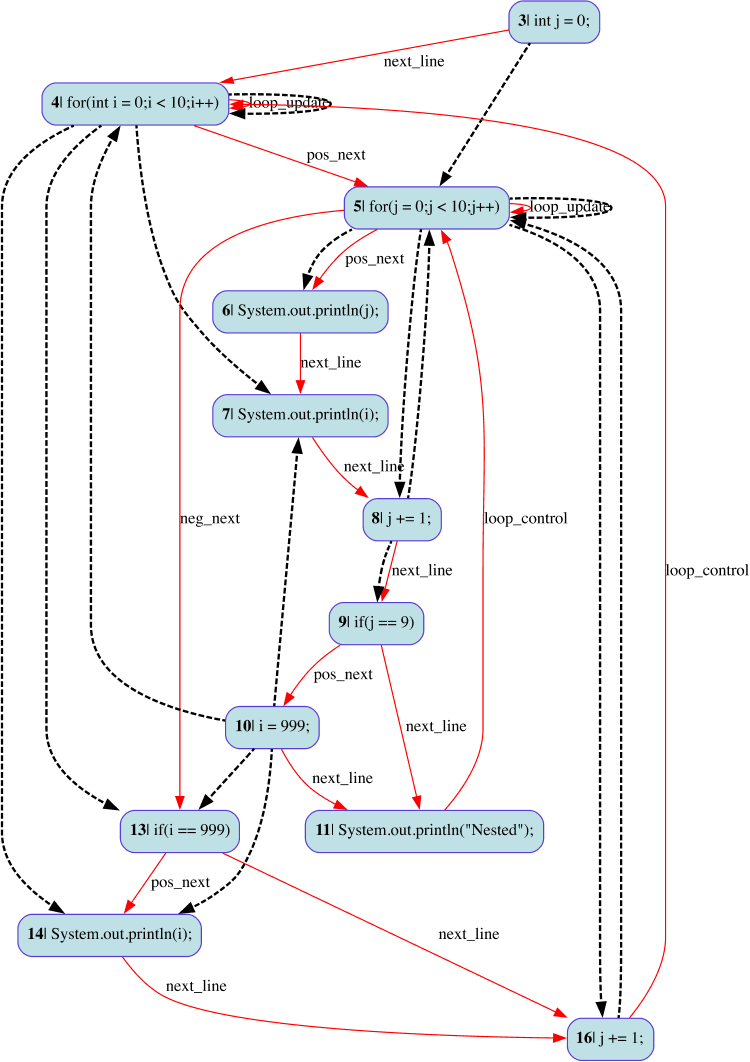}
    \caption{Example of CFG+DFG generated by \emph{COMEX} for a snippet that demonstrates the new and improved DFG implementation using RDA}
    \label{fig:dfg_rda}
\end{figure} 

The GraphCodeBERT module outputs a data flow graph as a collection of data-flow edges between pairs of AST leaf-nodes. This is different from COMEX that outputs data-flow edges between pairs of statements. Thus, to have a meaningful comparison, we first convert GraphCodeBERT's token-level output into a statement-level graph by mapping each token to its parent statement. The result of such a post-processing step can be seen in Fig. \ref{fig:dfg_gcb}. Upon further investigation, we observe that the GraphCodeBERT module provides a very simple data-flow implementation that works directly on a concrete syntax tree (CST). However, we observe that it misses out on key edges that a DFG should be able to capture. An example of this can be seen by contrasting the DFG obtained by GraphCodeBERT (Fig. \ref{fig:dfg_gcb}) against the corresponding output from \textit{COMEX} (Fig. \ref{fig:dfg_rda}) for the snippet shown in Listing \ref{lst:dataflow}. We highlight one issue with the help of a for-loop construct spanning Lines $5-12$. In this for-loop, the variable (\textit{j}) used in the condition at Line $5$, is also modified in the body of the loop at Line $8$. Hence, data flows from Line $8$ to Line $5$. 
In the case of \textit{COMEX}, there is a corresponding DFG edge from line numbers 8 to 5, while the same is missing in GraphCodeBERT (Fig. \ref{fig:dfg_gcb}). The GraphCodeBERT implementation also exhibits evident deficiencies in correctly handling scope and lacks inter-procedural analysis, among other shortcomings. 

Thus, we conclude that GraphCodeBERT's approximate analysis, while suitable for its approach, would not suffice our requirements. \textit{COMEX}, on the other hand, facilitates the generation of a statement-level DFG using RDA\footnote{Reaching Definition Analysis is a pivotal technique in data flow analysis \citep{aho2007compilers}. Its primary objective is to determine the set of definitions that have the potential to impact the value of a variable at a specific program point.} and also implements a more refined analysis of object behaviour like field accesses and method invocations. 
Although an RDA-based analysis inherently incurs higher computational costs, 
as compared to non-RDA analysis (like GraphcodeBERT), we opt for \textit{COMEX} to ensure that we use more accurate code-view graphs. 

\begin{lstlisting}[label={lst:dataflow},caption={Code Snippet corresponding to Fig. \ref{fig:dfg_gcb} and Fig. \ref{fig:dfg_rda}}, firstnumber=3,language=iJava]
int j = 0;
for (int i = 0; i < 10; i++) {
  for (j = 0; j < 10; j++) {
    System.out.println(j);
    System.out.println(i);
    j += 1;
    if (j == 9)
      i = 999;
    System.out.println("Nested");
  }
  if (i == 999) {
    System.out.println(i);
  }
  j += 1;
}
\end{lstlisting}
\begin{lstlisting}[label={lst:last_def},caption={Code Snippet corresponding to Fig. \ref{fig:last_def}}, firstnumber=2,language=iJava]
public int executeRemaining() throws SQLException {
  if (currentBatchSize > 0) {
      totalRecordsProcessed += currentBatchSize;
      preparedStatement = 0;
      currentBatchSize = 0;
  }
  // Avoid reuse, signal that this was cleanly closed.
  preparedStatement = null;
  LOG.info(String.format("Processed %d %s records", totalRecordsProcessed, recordType));
  return totalRecordsProcessed;
}
\end{lstlisting}
\begin{figure}
\centering
\hfil
{\includegraphics[width=0.9\textwidth]{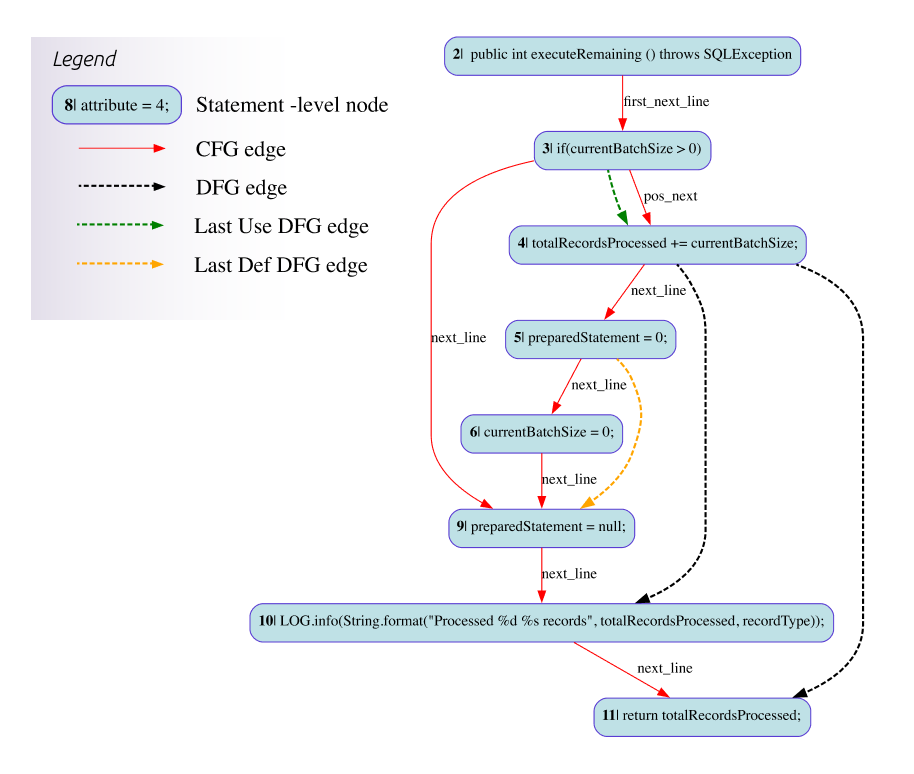}}
\caption{Data-flow graph illustrating last definition and last usage based data-flow edges}
\label{fig:last_def}
\end{figure}

\subsection{\textit{COMEX} Customizations used in \textit{CodeSAM}}

In addition to generating the typical code-views like AST, CFG and DFG, \textit{COMEX} also provides options to further customize these views, which proves extremely useful when integrating \textit{COMEX} into \textit{CodeSAM}. In particular, we exhaustively experiment with the ``Last Def" and ``Last Use" customization options that are available for the data-flow graph, dedicating separate results (with these options enabled) in each table in Section \ref{sec:experiments}. We notice that the ``Last Def" and ``Last Use" options ensure a more comprehensive coverage of variable dependencies for those code snippets that refer to many global variables and functions that are out-of-context or undefined. We depict this with the help of an example in  Fig \ref{fig:last_def}. As shown, the graph in Fig. \ref{fig:last_def} is a combination of the CFG, DFG, Last Use and Last Def edges for the code snippet in Listing \ref{lst:last_def}. 
The ``Last Use" relationship establishes links between the current usage of a variable and the program point where it was most recently read. 
Thus, the green edge from Line $3$ to Line $4$ is a ``Last Use" edge. This allows us to track the \textit{global} variable ``currentBatchSize" which would have otherwise been ignored by an RDA-based analysis due to its missing declaration.
With a similar motivation, we also enable the ``Last Def" relationship, which introduces extra edges that connect re-definitions of variables, allowing for the inclusion of edges between variable declarations and their corresponding definitions. The yellow edge from Line $5$ to Line $9$ in Fig. \ref{fig:last_def} is a ``Last Def" edge. Just as before, the ``preparedStatement" variable has not been declared in the scope of the provided code snippet. However, there exists a definition on Line $5$ and hence the closest viable definition is denoted as a data flow edge. Like ``Last Use", the ``Last Def" data-flow relationships thus prove particularly valuable in method-level snippets that heavily rely on global variables, which are not explicitly defined within the method body. 

In the following section, equipped with a flexible framework like \textit{CodeSAM} and a customizable code-view tool like COMEX, we detail our exhaustive experimental results spanning multiple code-view combinations across three common software engineering tasks.


\section{System, Experiments and Results}
\label{sec:experiments}
\begin{figure}
    \centering
    \includegraphics[width=0.7\textwidth]{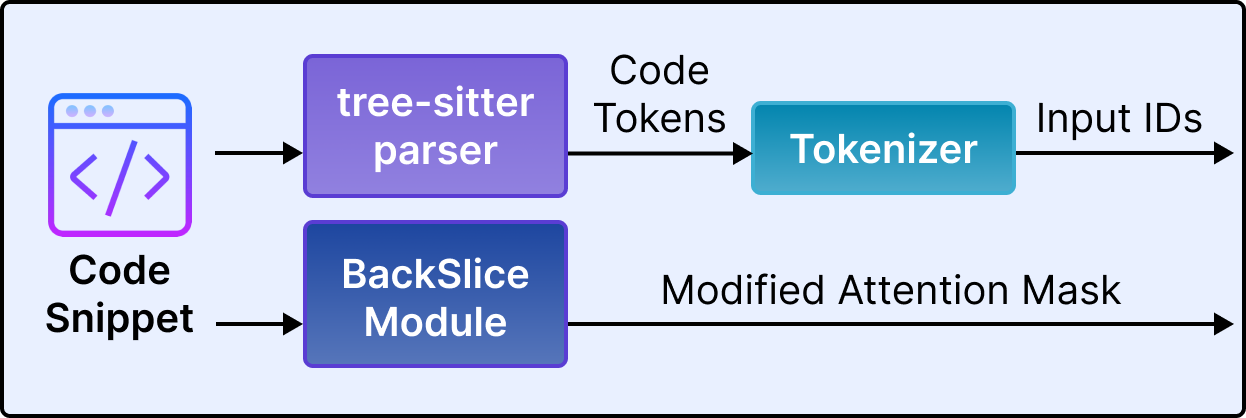}
    \caption{Input Pre-processor component of the experiment pipelines}
    \label{fig:input_preprocessor}
\end{figure}   
\begin{figure}
    \centering
    \includegraphics[width=0.7\textwidth]{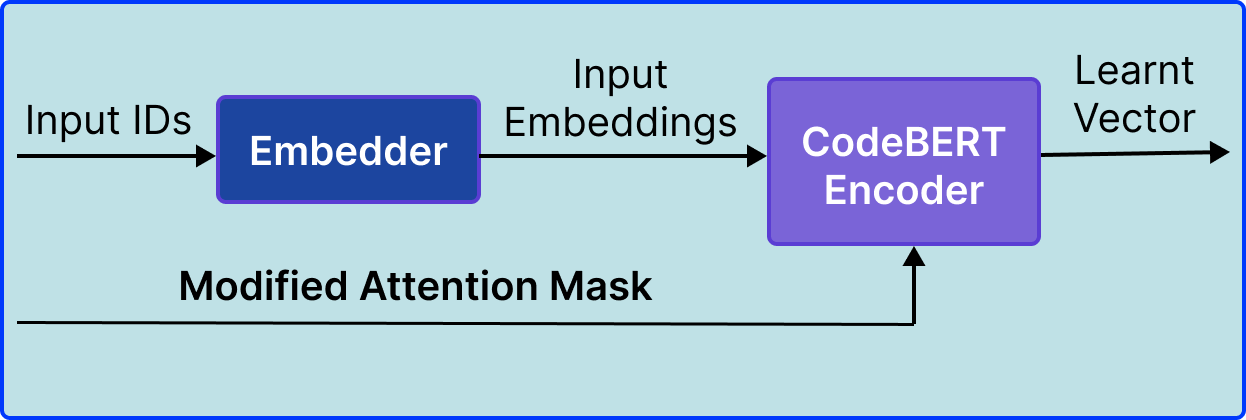}
    \caption{MCV Encoder component of the experiment pipelines}
    \label{fig:mcv_encoder}
\end{figure} 

To study the effectiveness of our approach, we run our end-to-end system (Fig. \ref{fig:system}), and feed different attention masks into the pre-trained transformer model CodeBERT \citep{https://doi.org/10.48550/arxiv.2002.08155}. CodeBERT is a bimodal transformer for programming and natural languages. It has been extensively trained on the considerably large ($\sim2.3$ million data points) \emph{CodeSearchNet} corpus \citep{codesearchnet} using masked language modelling (MLM) and replaced token detection \citep{feng2020codebert}. In our experiments, rather than pre-training a deep learning language model from scratch (which is compute-intensive), we \emph{fine-tune} CodeBERT with the attention mask $A'$ on the downstream tasks of semantic code search, program classification and code clone detection.
It is important to note that most transformer models (like CodeBERT) have multiple layers in their architecture, where each layer has its own self-attention operation. Hence, we can choose to apply $A'$ to every layer or selectively apply it to certain layers. In this work we experiment with two options - (a) \textbf{All Layer Masking}: Applying $A'$ to every layer and (b) \textbf{Alternate Layer Masking}: Applying $A'$ to every alternate layer in the model. Additionally, in certain code snippets, we observe that the masking percentage is exceedingly high (say $> 90\%$). This is because the code snippet refers to a lot of out-of-context information like undefined or global variables, or a myriad of method invocations on undefined objects. In such code snippets, $A'$ is very sparse, masking out a lot of information. In these scenarios, it is prudent to avoid masking in the model and instead let every token attend to all other tokens. Hence, we also perform some ablations with different masking limits ($70\%, 80\%, 90\%$). 


Using the above techniques, we additionally experiment with different code-view combinations and model ensembles to determine what works best for each task. In what follows, we first elaborate on some common architectural components and then explain our results for the downstream applications of semantic code search, program classification and code clone detection.

\subsection{Input Preprocessor and MCV Encoder}
Before going into the task-specific architectures, we first describe in detail two major components used in all of our pipelines - (a) Input Preprocessor (Fig. \ref{fig:input_preprocessor}) and (b) MCV Encoder (Fig. \ref{fig:mcv_encoder}). As shown in Fig. \ref{fig:input_preprocessor}, the Input Preprocessor is responsible for code snippet tokenization and attention mask generation ($A'$). For code snippet tokenization, we run the \textit{tree-sitter} parser on the code snippet and collect a list of leaf tokens as the input. This list is then further tokenized using CodeBERT's tokenizer resulting in a list of Input IDs. For the attention mask creation, we run Algorithm \ref{alg:backwardslice} on the input code snippet and generate $A'$. Once the code is tokenized and $A'$ is generated, we feed this to MCV Encoder (Fig. \ref{fig:mcv_encoder}) to generate a semantic vector representation of the code snippet. Based on the task, these vector representations are used to solve the different problems of search, classification and clone detection. In what follows, we describe in depth the pipeline for each task and our associated experimental results.

\subsection{Semantic Code Search}


\begin{figure*}
\centering
\includegraphics[width=\textwidth]{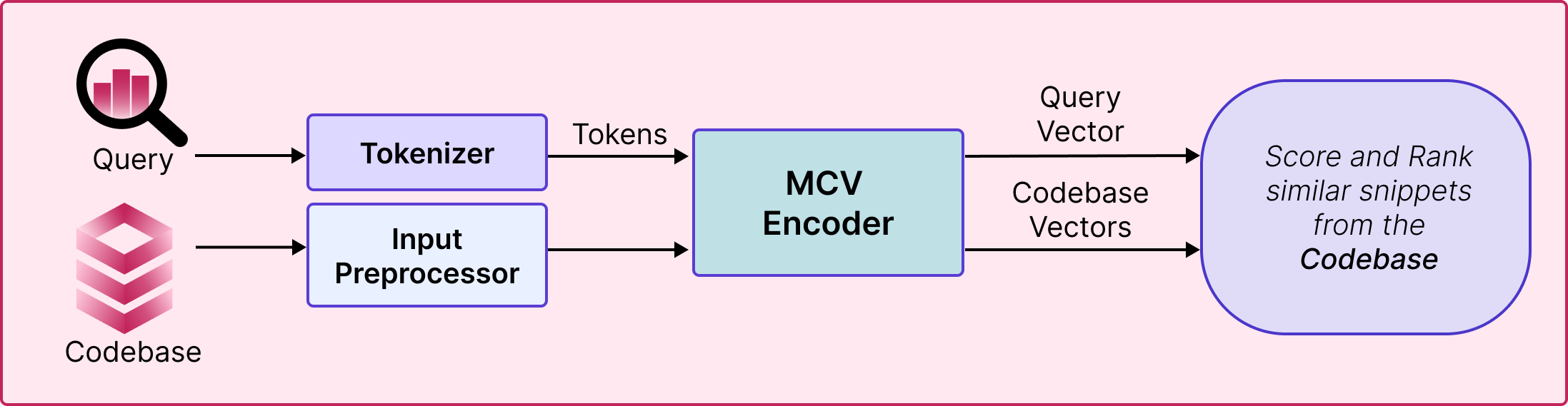}
\label{fig:semantic_code_search}
\caption{Pipeline Diagram for Semantic Code Search. (Input Preprocessor and MCV Encoder detailed in Fig. \ref{fig:input_preprocessor} and Fig. \ref{fig:mcv_encoder})}
\end{figure*}

Semantic code search is the task of retrieving a relevant code snippet from a code-base, given a natural language query. As software systems increasingly embrace open-source models, developers now have greater access to a vast array of source code. This accessibility allows developers to search for and utilize code snippets that can address similar (if not identical) tasks \citep{Di_Grazia_2023}, thereby improving productivity and reducing development time. By solving the problem of semantic code search, we aim to develop techniques that aid in this search process. It is important to note that code search has two modalities - natural language and code. Code, unlike natural language, is structured data and has unambiguous semantics. This nature of code poses a problem, as traditional search engines are trained to capture semantics from unstructured data (like text) and are hence not effective at code retrieval tasks \citep{10.1145/3196398.3196425}. To solve this problem, most semantic code search related works harness not only the natural language query but also the unique code structure to effectively retrieve the most relevant results.


\begin{lstlisting}[float=h, label={fig:codesearch:code_snippet},caption={A sample Java data point in the \textit{CodeSearchNet} dataset. The text above the function definition is the documentation corresponding to the code snippet below it.}, firstnumber=1,language=iJava]
/**
 * Add servlet names for the filter.
 * @param servletNames the servlet names to add
 */
public void addServletNames(String... servletNames) {
    Assert.notNull(servletNames, "ServletNames must not be null");
    this.servletNames.addAll(Arrays.asList(servletNames));
}
\end{lstlisting}

\subsubsection{Dataset}
We evaluate our performance on this task using the \textit{CodeSearchNet} Java corpus. This dataset was created by extracting pairs of documentation and code implementation from thousands of open-sourced GitHub repositories.  We present an example from \textit{CodeSearchNet} in Listing \ref{fig:codesearch:code_snippet}. As seen, the data point has a method-level summary along with a code implementation. Unfortunately, some data points in \textit{CodeSearchNet} are of poor quality. Thus, we reuse the modified version of this dataset as proposed by  \citep{guo2020graphcodebert}, wherein low-quality queries are removed using handcrafted rules. These handcrafted rules target those data points with excessively short documentation or code implementation.
The original dataset setting \citep{husain2020codesearchnet} restricts the number of candidate code snippets, during inference, to $1000$. This limitation does not reflect the true complexity of this task. Therefore, we follow \citep{ guo2020graphcodebert} and expand the total number of candidate code snippets to the entire code base. As shown in Table \ref{tab:code_search_data_stats}, the training dataset contains $164,923$ pairs of code snippets and corresponding documentation. Whereas, the test and the development datasets contain only $5,183$ and $10,955$ natural language queries (documentation) that are used to retrieve code from a code base with $40,347$ code snippets.

\begin{table}[h]
\centering
    \begin{subtable}[h]{0.45\textwidth}
        \centering
        \begin{tabular}{c|c}
            \textbf{Parameter} & \textbf{Value} \\
            \hline
            Batch Size & 128 \\
            Learning Rate & $2\times 10^{-5}$ \\
            Natural Language/ Code Length & 512 \\
            Epochs & 20
        \end{tabular}
    \caption{Experimental Setup for Semantic Code Search}
    \label{tab:codesearch_config}
    \end{subtable}
    \hspace{0.1\textwidth}
    \begin{subtable}[h]{0.45\textwidth}
        \centering
        \begin{tabular}{c|c}
            \textbf{Split} & \textbf{Count} \\
            \hline
            Train & 164,923 \\
            Dev & 5,183 \\
            Test & 10,955 \\
            Candidates   & 40,347
        \end{tabular}
    \caption{\textit{CodeSearchNet} Java Statistics}
    \label{tab:code_search_data_stats}
\end{subtable}
 \caption{Code Search hyper-parameters and \textit{CodeSearchNet} Statistics}
\end{table}


\subsubsection{Fine-Tuning Methodology} To adapt MCV Encoder for code search, we first perform fine-tuning with a contrastive loss function and then perform inference on the test set using the entire codebase.

\textbf{Fine-tuning}: The fine-tuning process in code search involves training the model to correctly identify matching pairs of code snippets and natural language descriptions. For this, we use a contrastive loss function, wherein, for every mini-batch, we obtain one set of embeddings for each natural language query and another set for each code snippet. Then we compute the dot product between these two sets of vectors, trying to maximize the dot product between matching pairs and minimizing the dot product between others. With these dot products, we calculate the cross-entropy loss and use backpropagation to train the model.

\textbf{Inference}: During inference or testing, we have access to a natural language query and the entire codebase of code snippets. To reduce inference time, we pre-compute and store the vector embeddings for each code snippet in the codebase. Then, for each natural language query we calculate the dot product between the natural language query vector and the set of pre-computed codebase vectors. These dot product results are then sorted, and the rank of the correct (``ground truth") snippet within this list is used to compute the Reciprocal Rank (Section \ref{MRR}).

\subsubsection{Metrics}
\label{MRR}

Semantic Code Search results are generally evaluated using the Mean Reciprocal Rank (MRR). The formula for MRR is depicted in Equation \ref{eqn:mrr_eqn}, where $rank_i$ is the actual position of the ground truth snippet in the list of sorted dot products for the $i^{th}$ natural language query. 

\begin{equation}
MRR = \frac{1}{|Q|}\sum_{i=1}^{|Q|}\frac{1}{rank_i}    
\label{eqn:mrr_eqn}
\end{equation}

The highest achievable MRR is $1$, where the ground truth is always at the $1^{st}$ position for all natural language queries.

\subsubsection{Results}

We fine-tune the underlying CodeBERT model (within MCV Encoder) using various configurations including $3$ code-views ($\mathbb{T}_{A}$, $\mathbb{T}_{D}$, $\mathbb{T}_{A+D}$), $3$ masking limits ($70\%$, $80\%$ and $90\%$), the addition/absence of Last Def and Last Use edges, and the choice of applying $A'$ to all layers or every alternate layer. Our generalizable approach and framework allows us to ensemble and choose from different code-view models that work best for the task. This kind of flexibility is notably absent in GraphCodeBERT as it is designed to incorporate only the data-flow graph. We also compare our results against other popular pre-trained models like RoBERTa \citep{liu2019roberta} and RoBERTa (code). RoBERTa  uses the MLM objective to pre-train on a text corpus whereas RoBERTa (code) uses the MLM objective to pre-train on a code corpus.

\begin{table}[h]
    \renewcommand{\arraystretch}{1.5}
    \centering
    \begin{tabular}{*{7}{|c}|}
    \hline
         & \multicolumn{6}{c|}{\textbf{Masking Limit Percentage}} \\
         \hline
            \textbf{Model} &  \multicolumn{3}{c|}{Vanilla} & \multicolumn{3}{c|}{Last Def \& Last Use} \\
            \hline
            & 70 & 80 & 90 & 70 & 80 & 90 \\ 
        \hline
            CodeBERT-$\mathbb{T}_{A}$	& 0.693 & \textbf{0.696} & 0.694 &
  \multicolumn{3}{c|}{N/A} \\
        \hline
            CodeBERT-$\mathbb{T}_{D}$ & 0.693 & 0.691 &	0.693 &	\textbf{0.695} & \textbf{0.695} &	0.691 \\
        \hline
            CodeBERT-$\mathbb{T}_{A+D}$ &0.696 &	0.697 &	0.694 &	0.697 &	\textbf{0.698} &	\textbf{0.698} \\
        \hline
            RoBERTa & \multicolumn{6}{c|}{0.599} \\
        \hline
            RoBERTa (code) & \multicolumn{6}{c|}{0.620} \\
        \hline
            CodeBERT & \multicolumn{6}{c|}{0.676} \\
        \hline
            GraphCodeBERT & \multicolumn{6}{c|}{0.691} \\
        \hline

    \end{tabular}
    \caption{Model performance (all layers masked) on Semantic Code Search as measured by MRR. }
    \label{tab:semantic_search_results}
\end{table}
\begin{table} [h]
    \renewcommand{\arraystretch}{1.5}
    \centering
    \begin{tabular}{|c|c|c|c|c|c|}
    \hline
    \textbf{\thead{Model}}   &  \textbf{\thead{Ensemble}} & \textbf{\thead{Masking \\ Limit \%}} & \textbf{\thead{Last Def \\  \& \\ Last Use}} & \textbf{\thead{MRR}} & \textbf{\thead{Ensemble \\MRR}} \\
    \hline
                        & CodeBERT-$\mathbb{T}_{A}$     &  80   &  No  & 0.696 &  \\
    CodeBERT-Ensemble   & CodeBERT-$\mathbb{T}_{D}$     &  70   &  No  & 0.695 & \textbf{0.710} \\
                        & CodeBERT-$\mathbb{T}_{A+D}$ &  80   &  Yes  & 0.698 &  \\ 
    \hline
    \end{tabular}
    \caption{CodeBERT ensemble performance (all layers masked) on Semantic Code Search as measured by MRR.}
    \label{tab:search_ensemble_results}
\end{table}

\textbf{All Layer Masking Results}: In the first set of experiments, we apply $A'$ to every self-attention layer and record the observed MRR scores in Table \ref{tab:semantic_search_results}. As shown, we observe that the best MRR scores for the AST code-view (CodeBERT-$\mathbb{T}_{A}$) is $0.696$, for the DFG code-view (CodeBERT-$\mathbb{T}_{D}$) is $0.695$ and for the AST+DFG code-view (CodeBERT-$\mathbb{T}_{A+D}$) is $0.698$. All three models surpass GraphCodeBERT which has an MRR score of $0.691$. Note that unlike GraphCodeBERT, we do not perform any extended pre-training. We achieve these results with a relatively compute-light fine-tuning. We also experiment with ensembles of the models and record their results in Table \ref{tab:search_ensemble_results}. In the ensemble, we take the best performing model from each code-view and average out the predictions to arrive at the final answer. The final ensemble MRR score is $0.710$ which is $2.75\%$ more than GraphCodeBERT.

\begin{table}[h]
    \renewcommand{\arraystretch}{1.5}
    \centering
    \begin{tabular}{*{7}{|c}|}
    \hline
         & \multicolumn{6}{c|}{\textbf{Masking Limit Percentage}} \\ \hline
            \textbf{Model} &  \multicolumn{3}{c|}{Vanilla} & \multicolumn{3}{c|}{Last Def \& Last Use} \\ \hline
            & 70 & 80 & 90 & 70 & 80 & 90 \\ 
        \hline
            CodeBERT-$\mathbb{T}_{A}$	& 0.684 & \textbf{0.689} & 0.680 &
 \multicolumn{3}{c|}{N/A} \\
        \hline
            CodeBERT-$\mathbb{T}_{D}$ & 0.680 & 0.687 &	0.688 & 
            0.680 & 0.683 &	\textbf{0.689} \\
        \hline
            CodeBERT-$\mathbb{T}_{A+D}$ & 0.678 & 0.687 & 0.686 &	
            \textbf{0.690} &	0.683 &	\textbf{0.690} \\
        \hline
            CodeBERT & \multicolumn{6}{c|}{0.676} \\
        \hline
            GraphCodeBERT & \multicolumn{6}{c|}{\textbf{0.691}} \\
        \hline
    \end{tabular}
    \caption{Model performance (alternate layers masked) on Semantic Code Search as measured by MRR.}
    \label{tab:semantic_search_results_alternate_masking}
\end{table}
\begin{table}[h]
    \renewcommand{\arraystretch}{1.5}
    \centering
    \begin{tabular}{|c|c|c|c|c|c|}
    \hline
    \textbf{\thead{Model}}   &  \textbf{\thead{Ensemble}} & \textbf{\thead{Masking \\ Limit \%}} & \textbf{\thead{Last Def \\  \& \\ Last Use}} & \textbf{\thead{MRR}} & \textbf{\thead{Ensemble MRR}} \\
    \hline
                        & CodeBERT-$\mathbb{T}_{A}$     &  80   &  No  & 0.689 &  \\
    CodeBERT-Ensemble   & CodeBERT-$\mathbb{T}_{D}$     &  80   &  Yes  & 0.689 & \textbf{0.701} \\
                        & CodeBERT-$\mathbb{T}_{A+D}$ &  90   &  Yes  & 0.690 &  \\ 
    \hline
    \end{tabular}
    \caption{CodeBERT ensemble performance (alternate layers masked) on Semantic Code Search as measured by MRR.}
    \label{tab:search_ensemble_results_alternate_masking}
\end{table}

\textbf{Alternate Layer Masking Results}: In the second set of experiments, we apply $A'$ on every alternate layer of MCV Encoder and record the results in Table \ref{tab:semantic_search_results_alternate_masking}.  Each individual code-view model is competitive but does not surpass GraphCodeBERT. Just as before, we also use an ensemble of these models and record an MRR score of $0.701$ swhich is $1.44\%$ higher than GraphCodeBERT but lower than the ensemble in Table \ref{tab:search_ensemble_results}. Hence, on the whole, for the task of code search, applying $A'$ to every layer seems to perform better than masking alternate layers.

\subsubsection{Experiment Hardware and Hyper-parameters}
To achieve the above results, we conduct our experiments with a batch size of $128$ and a learning rate of $2\times10^{-5}$. We train our model for $20$ epochs using $4$ CPUs, $196$ GB RAM and $4$ A100 GPUs. All other hyper-parameters are detailed in Table \ref{tab:codesearch_config}.
\newline
\newline
\fbox{\begin{minipage}{\hsize}
\textit{\textbf{CodeSearch Task Summary}} - Every individual code-view model that uses the all layer masking strategy surpasses GraphCodeBERT. But the highest MRR of \textbf{$0.71$} is achieved using an ensemble of the above models.
\end{minipage}}
\newline

\subsection{Clone Detection}

\begin{figure}[h]
\centering
\includegraphics[width=\textwidth]{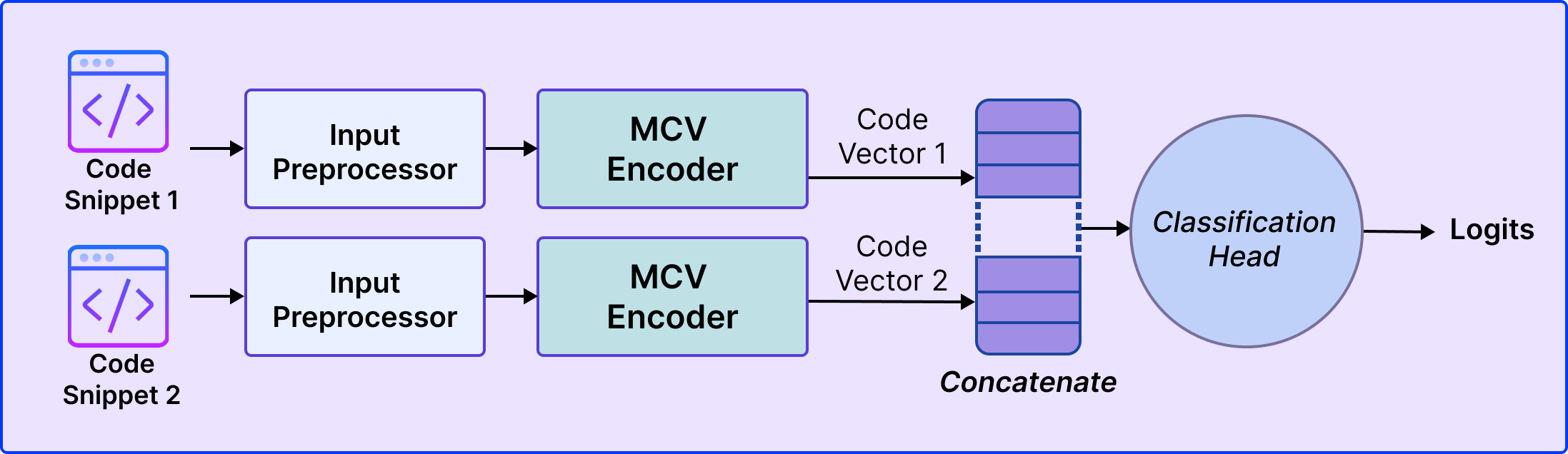}
\caption{Code Clone Detection (Input Preprocessor and MCV Encoder detailed in Fig. \ref{fig:input_preprocessor} and Fig. \ref{fig:mcv_encoder})}
\label{fig:code_clone_detection}
\end{figure}



As enterprise codebases grow in size, code maintenance eventually becomes a huge challenge. An important aspect of code maintenance is identifying and merging those code snippets that have very similar functionality. This merging allows for enhanced modularity, better consistency and higher code re-use across the entire codebase. However, manually identifying code segments that have similar functionalities can be very time consuming. Code clone detection provides a solution by predicting functionally similar code segments, irrespective of their syntactic differences. As a result, models trained on this task are particularly useful in identifying redundancies in a codebase, thereby reducing the cost of software maintenance. Code clone detection also helps in identifying plagiarism and copyright infringement in software systems \citep{Gold2021}.

\subsubsection{Dataset}



\begin{lstlisting}[float=h, label={fig:clone:code_snippet_1},caption={An example of code clones that perform the same function but differ in syntax and meaning significantly. Approach A in the listing computes the factorial of a number iteratively, while Approach B does it recursively.}, firstnumber=1,language=iJava]
// Approach A - An iterative approach
public int factorial(int number) {
    int result = 1;
    for (int i = 2; i <= number; i++) {
        result *= i;
    }
  return result;
}

// Approach B - A recursive approach
public int calculateFactorial(int number) {
  if (number <= 1) {
      return 1;
  } else {
      return number * calculateFactorial(number - 1);
  }
}
\end{lstlisting}

We evaluate the performance on this task using a filtered version of the \textit{BigCloneBench} dataset \citep{Svajlenko2021} as followed in \citep{wang2020detecting}. Each datapoint is a pair of code snippets along with a label indicating if the two snippets are clones. It is important to note that each code snippet is a \emph{method level} Java function as shown in Listing \ref{fig:clone:code_snippet_1}. In total, there are $\sim 1.7$ million datapoints in the dataset. The split between train, validation and test is shown in Table \ref{tab:clone_detection_dataset_stats}. 


\begin{table}[h]
    \begin{subtable}{0.45\textwidth}
    \centering
    \begin{tabular}{c|c}
        \textbf{Split} & \textbf{Count} \\
        \hline
        Train & 901,028 \\
        Test & 415,416 \\
        Validation & 415,416 \\
        \hline
        Total & 1,732,040 \\
    \end{tabular}
    \caption{\textit{BigCloneBench} dataset split statistics.}
    \label{tab:clone_detection_dataset_stats}
    \end{subtable}\hspace{0.1\textwidth}
    \begin{subtable}{0.45\textwidth}
    \centering
    \begin{tabular}{c|c}
            \textbf{Parameter} & \textbf{Value} \\
            \hline
            Batch Size & 32 \\
            Learning Rate & $2 \times 10^{-5}$ \\
            Natural Language/ Code Length & 512 \\
            Epochs & 2
        \end{tabular}
    \caption{Configuration for Code Clone Detection experiments}
    \label{tab:clone_detection_config}
    \end{subtable}
\caption{\textit{BigCloneBench} statistics and Code Clone Detection  hyper-parameters}
\end{table}



\subsubsection{Fine-Tuning Methodology}
As seen in Fig. \ref{fig:code_clone_detection}, we fine-tune MCV Encoder on the clone detection task by running the model separately for each method in the code snippet pair. By doing so, we produce two vectors, which we then concatenate and pass to a classification head (multi-layer perceptron). The classification head outputs a probability which along with the ground truth is used to calculate the binary cross entropy loss. During the course of training, the model learns to predict $1$ for a true clone pair and $0$ otherwise. In the inference stage, the pair of snippets are passed through MCV Encoder and if the output probability is more than $0.5$, we deem the pair to be a clone.

\subsubsection{Results}
The results of the code clone detection task are evaluated using precision, recall and F1-scores. Just like code search, we run our experiments on various combinations of code-views, masking limits, presence/absence of Last Def and Last Use edges as well as different masking strategies. We also report the results of FA-AST-GNN \citep{wang2020detecting} and RoBERTa (code) alongside our results. FA-AST-GNN utilizes an augmented AST with explicit control and data flow edges for code clone detection.

\begin{table}
    \renewcommand{\arraystretch}{1.5}
    \centering
    \begin{tabular}{*{8}{|c}|}
    \hline
     \multirow{2}{*}{\textbf{Model}} & \multirow{2}{*}{\textbf{\thead{Masking \\ Limit \%}}} & \multicolumn{3}{c|}{\textbf{Vanilla}} & \multicolumn{3}{c|}{\textbf{Last Def \& Last Use}} \\
         & & \textbf{F1} & \textbf{Recall} & \textbf{Precision} & \textbf{F1} & \textbf{Recall} & \textbf{Precision}   \\
        \hline
                               &   70      &  94.0        & 94.6     & 93.3 & \multicolumn{3}{c|}{\multirow{3}{*}{N/A}} \\
        CodeBERT-$\mathbb{T}_{A}$                &   80      & \textbf{94.4}         & \textbf{94.6}         & \textbf{94.2} & \multicolumn{3}{c|}{} \\
               &   90      & 94.4   & 94.1     & 94.6  & \multicolumn{3}{c|}{} \\
        \hline
                               &   70      &   94.6       & 95.0     & 94.3 &  94.4        & 94.8     & 94.0 \\
        CodeBERT-$\mathbb{T}_{D}$               &   80      &   \textbf{95.0}       & \textbf{93.8}     & \textbf{96.3} & \textbf{94.4}  & \textbf{94.6}  & \textbf{94.2} \\
                       &  90   &  94.9        & 94.6     & 95.1 & 94.1 & 94.7     & 93.4 \\
        \hline
        
                       &   70      & \textbf{94.8}         & \textbf{94.2}         & \textbf{95.4} &  \textbf{95}    &  \textbf{94.2}    &    \textbf{95.7}  \\
        CodeBERT-$\mathbb{T}_{A+D}$  &  80   &  94.6        & 94.4     & 94.9 &  94.2       & 93.9     & 94.5 \\
                       &   90      &  94.6        & 94.6     & 94.6 &  94.6  & 94.9  & 94.4  \\
        \hline
            FA-AST-GNN  & N/A & \textbf{95.0} & 94.0 & \textbf{96.0}& \multicolumn{3}{c|}{N/A} \\
        \hline
            RoBERTa (code) & N/A & 93.5 & 92.2 & 94.9 & \multicolumn{3}{c|}{N/A} \\
        \hline
            CodeBERT & N/A & 94.0 & 93.4 & 94.7 & \multicolumn{3}{c|}{N/A} \\
        \hline
            GraphCodeBERT & N/A & \textbf{95.0} & \textbf{95.2} & \textbf{94.8} & \multicolumn{3}{c|}{N/A} \\
        \hline
    \end{tabular}
    \caption{Model Performance (alternate layers masked) on Code Clone Detection as measured by F1, Recall and Precision.}
    \label{tab:code_clone_results_alternate}
\end{table}

\begin{table}
    \centering
    \renewcommand{\arraystretch}{1.5}
    \begin{tabular}{*{7}{|c}|}
    \hline
    \textbf{\thead{Model}} &    \textbf{\thead{Ensemble}} & \textbf{\thead{Masking \\ Limit \%}} & \textbf{\thead{Last Def \\ \& Last Use}} & \textbf{\thead{F1}} & \textbf{Recall} & \textbf{Precision} \\
    \hline
                        & CodeBERT-$\mathbb{T}_{A}$   & 80 &  & & &    \\
    CodeBERT-Vanilla    & CodeBERT-$\mathbb{T}_{D}$   & 80 & No & \textbf{95.1} & \textbf{94.4} & \textbf{95.9}   \\
                        & CodeBERT-$\mathbb{T}_{A+D}$ & 70 &  &  &  &      \\
    \hline
                        & CodeBERT-$\mathbb{T}_{A}$   & 80 & & & &    \\
    CodeBERT-LDLU    & CodeBERT-$\mathbb{T}_{D}$   & 80 & Yes & 95.0 & 95.1 & 95.0 \\
                        & CodeBERT-$\mathbb{T}_{A+D}$ & 90 & &  &  &      \\
    \hline
                        & CodeBERT-$\mathbb{T}_{A}$   & 80 & No & & &    \\
    CodeBERT-Mixed    & CodeBERT-$\mathbb{T}_{D}$   & 80 & No & 95.0 & 94.4 & 95.6 \\
                        & CodeBERT-$\mathbb{T}_{A+D}$ & 70 & Yes &  &  &      \\

    \hline
    \end{tabular}
    \caption{Ensemble Performance (alternate layers masked) on Code Clone Detection as measured by F1, Recall and Precision}
    \label{tab:clone_detection_ensemble_results_alternate}
\end{table}

\textbf{Alternate Layer Masking Results}: We start by showing our experimental results using the alternate layer masking strategy in Table \ref{tab:code_clone_results_alternate}. As seen, the best F1 score for DFG and AST+DFG matches GraphCodeBERT with a score of $95$, however their respective precision scores of $96.3$ and $95.7$ surpass GraphCodeBERT's precision score of $94.8$. FA-AST-GNN also achives an equal F1-score of $95.0$ with a slightly higher recall ($94.0$) and a slightly lower precision ($96.0$) than our best result CodeBERT-$\mathbb{T}_{D}$.
As our framework allows for an ensemble, we also experiment with combinations of models as shown in Table \ref{tab:clone_detection_ensemble_results_alternate}. We try CodeBERT-Vanilla - an ensemble of the best vanilla models across code-views, CodeBERT-LDLU - an ensemble of the best last-def-last-use models across code-views and CodeBERT-Mixed - an ensemble of the best models across code-views (irrespective of the last-def-last-use flag). We note that all three ensembles - CodeBERT-LDLU, CodeBERT-Vanilla and CodeBERT-Mixed surpass GraphCodeBERT on F1 as well as Precision. 

\textbf{All Layer Masking Results}: We depict our results with the all layer masking strategy in Table \ref{tab:code_clone_results}. Among the individual code-view models, the data-flow model achieves the highest F1 score of $94.8$ which is lesser than what was achieved in the alternate masking strategy. However, among the ensembles shown in Table \ref{tab:clone_detection_ensemble_results}, we note that CodeBERT-LDLU has the highest F1 score of $95.2$ which is higher than GraphCodeBERT as well as any other model that we have trained for clone detection.
\begin{table}[h]
    \renewcommand{\arraystretch}{1.5}
    \centering
    \begin{tabular}{*{8}{|c}|}
    \hline
        \multirow{2}{*}{\textbf{Model}} & \multirow{2}{*}{\textbf{\thead{Masking \\ Limit \%}}} & \textbf{F1} & \textbf{Recall} & \textbf{Precision} & \textbf{F1} & \textbf{Recall} & \textbf{Precision}  \\
        & & \multicolumn{3}{c|}{\textbf{Vanilla}} & \multicolumn{3}{c|}{\textbf{Last Def \& Last Use}} \\
        \hline
        &                         70      &  94.0    & 94.6    & 93.4 & \multicolumn{3}{c|}{} \\
        CodeBERT-$\mathbb{T}_{A}$               &   80      & \textbf{94.4}     & \textbf{94.5}     & \textbf{94.2} & \multicolumn{3}{c|}{N/A} \\
                               &   90      & 93.7     & 93.7     & 93.7 & \multicolumn{3}{c|}{}\\
        \hline
        
                             &   70      &   93.8    & 95.0     & 92.7 &  94.4    &   94.9    &    94.0  \\
        CodeBERT-$\mathbb{T}_{D}$             &    80      &   \textbf{94.8}    & \textbf{94.4}   & \textbf{95.3} & \textbf{94.6}    &  \textbf{94.1}    &   \textbf{95.2} \\
        &  90   &  93.8    &   94.5    & 93.1 & 94.0    &   94.2    &    93.9\\
        \hline
        
                      &   70      &  94.1    & 94.4     & 93.9 &  93.9    &   94.1    &    93.7 \\
        CodeBERT-$\mathbb{T}_{A+D}$   &  80   &  \textbf{94.4}  & \textbf{94.9} & \textbf{93.9} &  93.0   &   93.2   &     92.8 \\
                      &   90      &  94.0    & 94.0     & 93.9 &  \textbf{94.7}    &   \textbf{94.7}    &    \textbf{94.7} \\
        \hline
            CodeBERT & N/A & 94.0 & 93.4 & 94.7 & \multicolumn{3}{c|}{N/A} \\
        \hline
            GraphCodeBERT & N/A & \textbf{95.0} & \textbf{95.2} & \textbf{94.8} & \multicolumn{3}{c|}{N/A} \\
        \hline
    \end{tabular}
    \caption{Model Performance (all layers masked) on Code Clone Detection as measured by F1, Recall and Precision.}
    \label{tab:code_clone_results}
\end{table}
\begin{table}[h]
    \centering
    \renewcommand{\arraystretch}{1.5}
    \begin{tabular}{*{7}{|c}|}
    \hline
    \textbf{\thead{Model}} &    \textbf{\thead{Ensemble}} & \textbf{\thead{Masking \\ Limit \%}} & \textbf{\thead{Last Def \\ \& Last Use}} & \textbf{\thead{F1}} & \textbf{Recall} & \textbf{Precision} \\
    \hline
                        & CodeBERT-$\mathbb{T}_{A}$   & 80 &  & & &    \\
    CodeBERT-Vanilla    & CodeBERT-$\mathbb{T}_{D}$   & 80 & No & 94.8 & 94.8 & 94.8    \\
                        & CodeBERT-$\mathbb{T}_{A+D}$ & 80 &  &  &  &      \\
    \hline
                        & CodeBERT-$\mathbb{T}_{A}$   & 80 & & & &    \\
    CodeBERT-LDLU    & CodeBERT-$\mathbb{T}_{D}$   & 80 & Yes & \textbf{95.2} & \textbf{95.0} & \textbf{95.4} \\
                        & CodeBERT-$\mathbb{T}_{A+D}$ & 90 & &  &  &      \\
    \hline
                        & CodeBERT-$\mathbb{T}_{A}$   & 80 & No & & &    \\
    CodeBERT-Mixed    & CodeBERT-$\mathbb{T}_{D}$   & 80 & No & 94.7 & 94.7 & 94.7 \\
                        & CodeBERT-$\mathbb{T}_{A+D}$ & 90 & Yes &  &  &      \\
    \hline 
    \end{tabular}
    \caption{Ensemble Performance (all layers masked) on Code Clone Detection as measured by F1, Recall and Precision}
    \label{tab:clone_detection_ensemble_results}
\end{table}

\subsubsection{Experiment Hardware and Hyper-parameters}
To obtain the above results, we perform our experiments using $2$ CPUs with $196$ GB RAM and $2$ V100 GPUs. We train our models for $2$ epochs with a batch size of $32$ and a learning rate of $2 \times 10 ^ {-5}$. Other hyper-parameters are detailed in Table \ref{tab:clone_detection_config}. 
\newline
\newline
\fbox{\begin{minipage}{\hsize}
\textbf{\textit{Clone Detection Task Summary}} - The AST and AST+DFG code-view models using the alternate layer masking strategy match GraphCodeBERT's F1 score and surpass GraphCodeBERT's precision. The CodeBERT-Vanilla ensemble (alternate layers masked) beats GraphCodeBERT on both F1 as well as Precision. But the highest F1 score is achieved by the CodeBERT-LDLU ensemble (all layers masked) that beats GraphCodeBERT and all other variations.
\end{minipage}}

\subsection{Program Classification}

\begin{figure}[h]
\centering
\includegraphics[width=\textwidth]{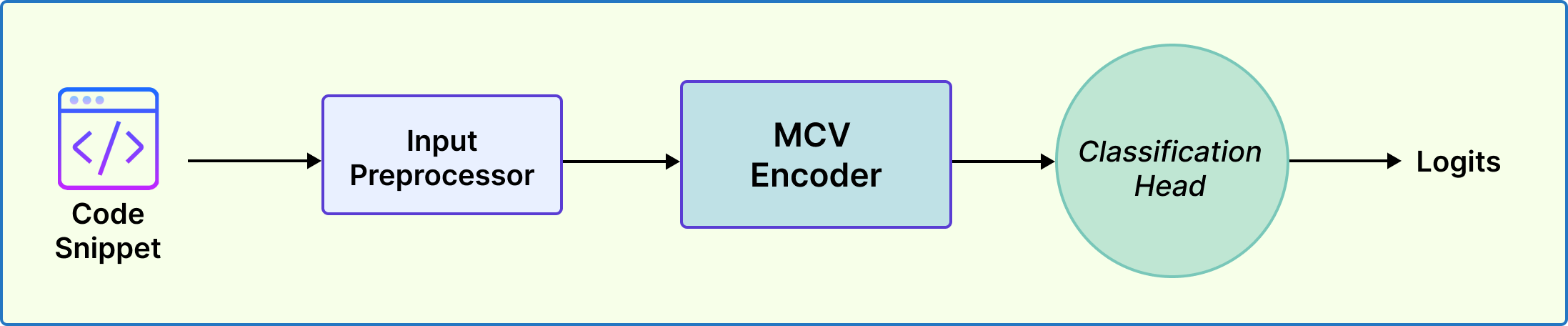}
\caption{Program Classification (Input Preprocessor and MCV Encoder detailed in Fig. \ref{fig:input_preprocessor} and Fig. \ref{fig:mcv_encoder})}
\label{fig:program_classification}
\end{figure}

Our third and final evaluation task is program classification. The goal of program classification is to categorize a given code snippet into different groups or classes based on the code snippet's functionality, code quality or algorithmic implementation. Using code classification techniques, we can categorize large codebases into similar groups/functionalities, allowing developers to easily navigate and manage these codebases. One common application of code classification is in software maintenance and re-engineering, where large legacy codebases need to be analyzed and restructured to meet changing business requirements \citep{BARCHI2021104075}. Program classification serves as a tool for pinpointing components of the codebase that require extensive refactoring or complete redesign \citep{kaur2016neural}. Moreover, it can aid in highlighting potential areas of risk due to bugs or vulnerabilities \citep{8614145}.

\subsubsection{Dataset}
We evaluate our performance on code classification using the Java \textit{CodeNet} corpus \citep{ibmcodenet}. The dataset consists of code snippets extracted from online judge websites. Each code snippet (or code submission) solves a particular problem statement with a unique problem ID. After assigning the relevant problem ID to each code submission, we end up with a single-label multi-class dataset, which can be modelled as a classification problem. Additionally, each snippet in this dataset is a \emph{file-level} snippet consisting of multiple methods and package imports. 
The dataset contains $75,004$ Java code snippets spread across $250$ classes. The split between train, validation and test is summarized in Table \ref{tab:classification_stats}.

\begin{table}[ht]
    \begin{subtable}{0.4\textwidth}
    \centering
    \begin{tabular}{c|c}
        \textbf{Split} &  \textbf{Count} \\
        \hline
        Train & 48,002 \\
        Test & 15,001 \\
        Validation & 12,001 \\
        \hline
        Total & 75,004
    \end{tabular}
    \caption{Code Classification Dataset Split}
    \label{tab:classification_stats}
    \end{subtable}
    \hspace{0.1\textwidth}
    \begin{subtable}{0.4\textwidth}
    \centering
    \begin{tabular}{c|c}
        \textbf{Parameter} & \textbf{Value} \\
        \hline
        Batch Size & 32 \\
        Learning Rate & $2 \times 10^{-5}$ \\ 
        Tokenizer & microsoft/codebert-base \\
        Code Length & 512 \\
    \end{tabular}
    \caption{Experiment Setup for Code Classification}
    \label{tab:classification_config}
    \end{subtable}
    \caption{Code Classification Statistics and Hyper-parameters}
\end{table}


\subsubsection{Fine-Tuning Methodology} As shown in Fig. \ref{fig:program_classification}, we fine-tune MCV Encoder on program classification using an additional classification head and then subsequently perform inference on the test set. More specifically, we attach a compute-light classification head (multi-layer perceptron) to the MCV Encoder that takes as input a vector embedding and outputs a $250$ label array (logits). After this, we apply the softmax operation to this array and compute the cross entropy loss against the ground truth, so that the model learns to predict the correct class. During the Inference stage, we leverage our trained model in Fig. \ref{fig:program_classification} to generate a probability distribution across the $250$ classes by using a softmax operation. We then assign the class with the highest probability to the code snippet.

\begin{table}
    \renewcommand{\arraystretch}{1.5}
    \centering
    \begin{tabular}{*{7}{|c}|}
    \hline
         & \multicolumn{6}{c|}{\textbf{Masking Limit Percentage}} \\
            \textbf{Model} &  \multicolumn{3}{c|}{Vanilla} & \multicolumn{3}{c|}{Last Def \& Last Use} \\
            & 70 & 80 & 90 & 70 & 80 & 90 \\ 
        \hline
            CodeBERT-$\mathbb{T}_{A}$	& \textbf{98.33} & 98.19 & 98.18 & \multicolumn{3}{c|}{N/A} \\
        \hline
            CodeBERT-$\mathbb{T}_{D}$ & \textbf{98.40} & 98.24 & 98.34 & \textbf{98.40} & 98.31 & 98.35 \\
        \hline
            CodeBERT-$\mathbb{T}_{A+D}$ & 98.24 & 98.26 & \textbf{98.33} & \textbf{98.29} & 98.23 & 98.28 \\
        \hline
            CodeBERT & \multicolumn{3}{c|}{ \textbf{98.41} } & \multicolumn{3}{c|}{N/A} \\
        \hline
    \end{tabular}
    \caption{Model Performance (alternate layers masked) on Program Classification as measured by macro-F1}
    \label{tab:classification_alternate_masking_results}
\end{table}

\begin{table}
    
    \centering
    \renewcommand{\arraystretch}{1.5}
    \begin{tabular}{*{5}{|c}|}
    \hline
    \textbf{\thead{Model}} &    \textbf{\thead{Ensemble}} & \textbf{\thead{Masking \\ Limit \%}} & \textbf{\thead{Last Def \\ \& Last Use}} & \textbf{\thead{F1-Score}} \\
    \hline
                        & CodeBERT-$\mathbb{T}_{A}$   & 70 & & \\
    CodeBERT-Vanilla    & CodeBERT-$\mathbb{T}_{D}$   & 70 & No & \textbf{98.51}  \\
                        & CodeBERT-$\mathbb{T}_{A+D}$ & 90 & & \\
    \hline
                                    & CodeBERT-$\mathbb{T}_{A}$   & 70 & & \\
    CodeBERT-LDLU   & CodeBERT-$\mathbb{T}_{D}$   & 70 & Yes & \textbf{98.51}  \\
                                    & CodeBERT-$\mathbb{T}_{A+D}$ & 70 & & \\
    \hline
                                & CodeBERT-$\mathbb{T}_{A}$   & 70 &  No & \\
    CodeBERT-Mixed   & CodeBERT-$\mathbb{T}_{D}$   & 70 & No & 98.44  \\
                                    & CodeBERT-$\mathbb{T}_{A+D}$ & 70 & Yes & \\
    \hline
    

        
    \end{tabular}
    \caption{ Ensemble Performance (alternate layers masked) on Program Classification as measured by macro-F1}
    \label{tab:classification_ensemble_results_wo_mask}
\end{table}



\subsubsection{Results} We evaluate our approach on program classification using the macro F1 score. Each data point in Java \textit{CodeNet} is a complete Java file.  GraphCodeBERT lacks support for file-level program analysis, unlike COMEX that accommodates both method-level and file-level analysis. Consequently, for this task, our comparison is restricted to CodeBERT.

\textbf{Alternate Layer Masking Results}:  Table \ref{tab:classification_alternate_masking_results} depicts the results for individual code-view models trained with the alternate masking strategy. As seen, the best performing code-view model is DFG achieving a macro F1 score of $98.40$, which is competitive with CodeBERT ($98.41$). As earlier, we experiment with three types of ensembles and note down our results in Table \ref{tab:classification_ensemble_results_wo_mask}. From Table \ref{tab:classification_ensemble_results_wo_mask}, we observe that CodeBERT-Vanilla and CodeBERT-LDLU ensembles achieve F1 scores of $98.51$ which is significantly higher than CodeBERT.


\begin{table}
    \renewcommand{\arraystretch}{1.5}
    \centering
    \begin{tabular}{*{7}{|c}|}
    \hline
         & \multicolumn{6}{c|}{\textbf{Masking Limit Percentage}} \\
            \textbf{Model} &  \multicolumn{3}{c|}{Vanilla} & \multicolumn{3}{c|}{Last Def \& Last Use} \\
            & 70 & 80 & 90 & 70 & 80 & 90 \\ 
        \hline
            CodeBERT-$\mathbb{T}_{A}$	& \textbf{98.22} & 98.18 & 97.99 & \multicolumn{3}{c|}{N/A} \\
        \hline
            CodeBERT-$\mathbb{T}_{D}$ & 98.13 &	98.11 &	\textbf{98.14}  &	\textbf{98.26} &	98.11 &	98.06 \\
        \hline
            CodeBERT-$\mathbb{T}_{A+D}$ & \textbf{98.14} & 98.07 & 98.05 & 98.05 & \textbf{98.08} & 98.04 \\
        \hline
    \end{tabular}
    \caption{Model Performance (all layers masked) on Program Classification as measured by macro-F1}
    \label{tab:classification_results}
\end{table}

\begin{table}
    \centering
    \renewcommand{\arraystretch}{1.5}
    \begin{tabular}{*{5}{|c}|}
    \hline
    \textbf{\thead{Model}} &    \textbf{\thead{Ensemble}} & \textbf{\thead{Masking \\ Limit \%}} & \textbf{\thead{Last Def \\ \& Last Use}} & \textbf{\thead{F1-Score}} \\
    \hline
                        & CodeBERT-$\mathbb{T}_{A}$   & 70 & & \\
    CodeBERT-Vanilla    & CodeBERT-$\mathbb{T}_{D}$   & 90 & No & 98.38  \\
                        & CodeBERT-$\mathbb{T}_{A+D}$ & 70 & & \\
    \hline
                                    & CodeBERT-$\mathbb{T}_{A}$ & 70 & & \\
    CodeBERT-LDLU   & CodeBERT-$\mathbb{T}_{D}$   & 70 & Yes & 98.35  \\
                                    & CodeBERT-$\mathbb{T}_{A+D}$ & 80 & & \\
    \hline
                                & CodeBERT-$\mathbb{T}_{A}$   & 70 &  No & \\
    CodeBERT-Mixed   & CodeBERT-$\mathbb{T}_{D}$   & 70 & Yes & 98.36  \\
                                    & CodeBERT-$\mathbb{T}_{A+D}$ & 70 & No & \\

    \hline
    
    

    
    \end{tabular}
    \caption{Ensemble Performance (all layers masked) on Program Classification as measured by macro-F1}
    \label{tab:classification_ensemble_results}
\end{table}

\textbf{All Layers Masking Results}: Table \ref{tab:classification_results} and Table \ref{tab:classification_ensemble_results} outline the results of program classification with the all layers masking strategy. From the observed results, it is evident that the alternate masking approach is much more effective for program classification. This is also clear from the fact that CodeBERT-$\mathbb{T_{D}}$ (in Table \ref{tab:classification_alternate_masking_results}) has a better F1 score than all three ensembles in Table \ref{tab:classification_ensemble_results}. As a result, we believe that the alternate layer masking strategy is better suited for program classification than masking every layer.




\subsubsection{Experiment Hardware and Hyper-parameters.}
In all our experiments, we use a batch size of $32$, a learning rate of $2 \times 10^{-5}$ and an Adam optimizer with a linear scheduler. We run our experiments on $4$ CPUs with $360$ GB RAM and $2$ V100 GPUs for a total of $2$ epochs. Additional hyper-parameters for the experiment are provided in Table \ref{tab:classification_config}.
\newline
\newline
\fbox{\begin{minipage}{\hsize}
\textit{\textbf{Program Classification Summary}} - The DFG code-view model using the alternate layer masking strategy is competitive with GraphCodeBERT's macro-F1 score. However, CodeBERT-Vanilla and CodeBERT-LDLU are two ensembles of the alternate masking strategy that beat GraphCodeBERT by a significant margin.
\end{minipage}}
\newline
\newline
\textbf{Final Summary}: Based on the results obtained across the three tasks, we conclude that there is merit in leveraging multiple code-views when learning effective representations of source code. In all three tasks, we see a visible improvement across metrics without the need for any extra pre-training.   



\section{Related Work}
\label{sec:related_work}
In this section, we go over the diverse literature available for source code representation learning. We categorize most of the approaches into four broad types and elaborate on relevant works in each section: 

\subsection{Token-based Representation}
\emph{Token-based Representation} approaches such as
\citep{movshovitz2013natural,allamanis2016convolutional} consider code as natural language tokens. They represent source code as a bag of words or a list of tokens, and then apply techniques like Latent Semantic Indexing (LSI) and Latent Dirichlet Allocation (LDA). 
In \citep{movshovitz2013natural}, the authors predict comments from Java source files of open source projects, using topic models and n-grams. 
\citep{allamanis2016convolutional} introduces an attention neural network that employs convolution on the code input tokens to detect local time-invariant and long-range topical attention features in a context-dependent way. They evaluate their methodology on the task of extreme summarization of source code, converting code snippets into short function-name-like summaries.  

\subsection{Structure-based Representation}
 Unlike natural languages, a lot of important information is contained in the structure of programming languages. These syntactic and semantic properties of source code can be extracted using structured or graphical representations like the AST \citep{zhang2019novel,wei2017supervised,white2016deep}, CFG \citep{zhao2018deepsim}, and DFG \citep{li2019improving}. Source code representation learning techniques of this type are grouped under \emph{Structure-based Representation} approaches. Rather than using the entire AST, \citep{zhang2019novel} presents a novel ASTNN approach that splits each large AST into a sequence of small statement trees, and encodes the statement trees to vectors by capturing the lexical and syntactical knowledge of statements. 
 \citep{wei2017supervised} presents a deep feature learning framework called CDLH that exploits the lexical and syntactical information extracted from an AST for fast computation of functional similarity between code fragments. 
 \citep{white2016deep} introduced a learning-based approach to the four types of code clone detection that harnesses information from both identifiers and structure through the AST. They evaluated their work and saw significant improvements over the traditional, structure based approaches and the prominent clone detection tool Deckard \citep{jiang2007deckard}. \citep{zhao2018deepsim} showed improvements over Deckard, CDLH \citep{wei2017supervised} and RtvNN \citep{white2016deep} by proposing a novel approach that encodes code control-flow and data-flow into a semantic matrix in which each element is a high-dimensional sparse binary feature vector. 
\emph{Combined Representation} or \textit{Multi Code-view} methods attempt to combine different structured representations. 
\citep{long2022multi} proposes a multi-view graph (MVG) program representation method that pays more attention to code semantics and simultaneously includes a DFG, a CFG and a read-write graph (RWG) as multiple views. These views are then combined and processed by a graph neural network (GNN) to obtain a comprehensive program representation which is evaluated in the context of the algorithm detection task using the POJ-104 and a custom and more challenging dataset called ALG-109. \citep{gao2023code} proposes \textit{SG-Trans}, a model for source code summarization that incorporates local and global code structural properties into the self-attention module of a transformer as inductive bias.

\subsection{Sequence-based Representation}
Most of the best results on a large variety of downstream tasks, are obtained from language models that are trained on both natural language and programming languages, drawing heavy inspiration from NLP techniques. We group these contributions \citep{feng2020codebert, nijkamp2022codegen, xu2022systematic} under \emph{Sequence-based Representation} approaches. CodeBERT \citep{feng2020codebert} is a pre-trained bimodal language model that learns both programming languages (PL) and natural language (NL). CodeBERT has a Transformer-based architecture and is pre-trained on NL-PL pairs along with unimodal code data. It is performant on both natural language code search and code documentation generation tasks. CodeGen, \citep{nijkamp2022codegen}, a large language model for code with multi-turn program synthesis, is trained on both NL and PL data. Their evaluation shows that the same intent provided in a multi-turn fashion significantly improves program synthesis over that provided as a single turn. \citep{xu2022systematic} proposes PolyCoder, based on GPT-2 architecture, and evaluates its results across 12 programming languages, while comparing and contrasting with some of the most popular large language models like Codex, GPT-J, GPT-Neo, GPT-NeoX-20B, and CodeParrot. 

\subsection{Sequence-from-structure-based Representation}
\emph{Sequence-from-structure-based Representation} approaches take one or more highly structured code-views (usually graphs like the AST, CFG, or DFG), and extract information as sequences by flattening them \citep{alon2019code2vec, swarna2024impact, guo2020graphcodebert}. Code2Vec is one of the pioneering works in the domain of source code representation learning where sequences were extracted from the AST of a code snippet by selecting random leaf-to-leaf paths. Using a simple neural network with a single attention layer for the model, the method was tested on the method naming task and showed a significant improvement in performance. The Mocktail approach \citep{swarna2024impact} extends the path-extracting methodology of code2vec \citep{alon2019code2vec} but parallelly extracts and combines paths from the AST, PDG and CFG. The approach is evaluated on the method naming task. \citep{fang2020functional} proposes a novel joint code representation that applies fusion embedding techniques to learn hidden syntactic and semantic features of source code. The paper provides interesting insights into results obtained from word embedding and graph embedding techniques from a combined code graph. Finally, the GraphCodeBERT approach \citep{guo2020graphcodebert}, achieving excellent performance on multiple tasks such as code clone detection, semantic code search, code translation and code refinement, appends data-flow sequences to the input sequence of code tokens. The model is an extension of the CodeBERT \citep{feng2020codebert} model and is pretrained on an additional task that includes data-flow information. Closely related to CodeSAM, \citep{du2023pre} proposes a directed, multiple-label code graph representation named Semantic Flow Graph (SFG) which compactly and adequately captures code semantics and achieves state-of-the-art performance in bug localization by combining it with a novel contrastive learning technique. An interesting fused representation for vulnerability detection using carefully extended neural network models to extract vulnerability-indicative semantic features from the token sequence, attributed control flow graph (ACFG), and AST representations of a function is presented in \citep{tian2023learning}. \citep{tian2024enhancing} beats the state-of-the-art at vulnerability detection by using a transformer-style encoder to aggregate the long-range contextual semantics of AST sub-trees into a vulnerability-specific vector to represent the target code fragment.


\section{Threats to Validity}
\label{sec:threats}

In this section, we discuss the potential threats to the validity of our study. 

\emph{Internal threats} - In our study, we deliberately chose a static analysis tool that operates directly on source code, enabling us to analyze even non-compilable data points. As a result, the code-views generated by the selected tool provide an approximation of control-flow and data-flow information. Therefore, findings derived from these code-views might introduce inaccuracies in our pipeline. In Section \ref{sec:experiments}, for each task and dataset we have clearly outlined the techniques and modifications incorporated during data pre-processing . While we follow established practices from existing literature, 
pre-processing decisions can introduce biases or limitations that may impact the study's outcomes.
Another potential threat to the validity of our study could be the bias or over-fitting of our models. As we use different models for different tasks, the hyper-parameters of the model need to be appropriately tuned. Even though our models are exhaustively validated, over-fitting can occur if the data is not diverse or due to a lack of sufficient regularization strategies.

\emph{External threats} - Our work has been evaluated on datasets which are widely used as benchmarks for the respective SE tasks. This enables us to demonstrate the effectiveness of our models compared to existing baselines. However, the generalizability of our results is thus subject to how accurately these datasets represent the complexity of real-world code corpora. \textit{BigCloneBench}, a widely accepted dataset, routinely used as a benchmark for clone detection in Java has been recently criticized for being imbalanced and  biased due to the way it was constructed \citep{9978255}. 
Due to a lack of a more comprehensive and well-known benchmark that fits our use case, our code clone detection evaluations are limited to BCB, in line with GraphCodeBERT which we draw comparisons against.  



\section{Conclusion and Future Work}
\label{sec:conclusion}
In this work, we introduced a framework that allows researchers to infuse custom code-views into self-attention based transformer models. We propose and evaluate a novel approach to infusing information from various highly structured code-views into language models to further boost their performance. To implement this pipeline, we investigated and carefully selected the the most suitable tool that allows generating outputs tailored to our specific requirements from amongst currently available open-source options to generate code-views. We then observe the impact on performance when using various customizes combinations of code-views through this approach on the downstream SE tasks of semantic code search, code clone detection, and program classification. 

Our work has several important applications and extensions. The \textit{CodeSAM} mechanism proposed in this work provides an exciting start-point for various language model explorations. For example, a logical extension to explore would be to pre-train a language model on a large dataset (like \emph{CodeSearchNet}) using code-view inspired attention masks (like $A'$) or
even code-view inspired novel pre-training objectives like \textit{ast-edge prediction} and \textit{next-control-statement prediction}. In
addition, exploring the generalisability of the approach to more diverse SE tasks like code translation, code summarization, code optimization, and code visualization, and understanding the influence of a code-view on a specific downstream task is another line of study to research. Since the framework is flexible and highly customizable, researchers can also look into cross-language code similarity learning and its subsequent applications.

\bibliography{sn-bibliography}

\end{document}